\documentclass[aps,twocolumn,%showpacs,preprintnumbers,
 superscriptaddress,amsmath,amssymb,nofootinbib]{revtex4-2}
\usepackage{amsfonts}
\usepackage{amsmath}
\usepackage{bbm}
\usepackage{amssymb}
\usepackage{graphicx}%
\usepackage{footmisc}
\usepackage{bm}
\usepackage{braket}
\usepackage{CJK}
\usepackage{hyperref}
\usepackage[capitalise]{cleveref}
\usepackage{longtable,booktabs}
\usepackage{subfigure}
\usepackage{booktabs}  % 三线表支持
\usepackage{pifont}
\newcommand{\cmark}{\ding{51}} % 勾号定义
\newcommand{\ctimes}{\ding{55}} % 叉号定义

\newcommand{\abs}[1]{\lvert#1\rvert}
\newcommand{\norm}[1]{\lVert#1\rVert}
\newcommand{\Tr}{\text{Tr}}

\begin{document}
\begin{CJK*}{UTF8}{gbsn}

\title{The trade-off between diagonal and off-diagonal elements in the eigenstate thermalization hypothesis }

\author{Zhiqiang Huang (黄志强)}
\email{hzq@wipm.ac.cn}
\affiliation{State Key Laboratory of Magnetic Resonance and Atomic and Molecular Physics, Innovation Academy for Precision Measurement Science and Technology, Chinese Academy of Sciences, Wuhan 430071, China}

\date{\today}

\begin{abstract}
    To bypass the reliance on local observables in verifying the eigenstate thermalization hypothesis (ETH), we introduce an observable-independent measure of distinguishability based on the variance of a rescaled local operator. We establish a universal trade-off relation between the diagonal and off-diagonal elements of this measure, rigorously connecting it to eigenstate typicality and spatially averaged observables. This trade-off reveals that exponential growth in the number of off-diagonal terms enforces their suppression, indirectly constraining diagonal deviations. Numerical simulations on a one-dimensional Ising spin chain with tunable transverse and longitudinal fields demonstrate stark contrasts between integrable and non-integrable regimes: While off-diagonal elements are universally suppressed with system size, diagonal suppression fails in integrable systems due to the absence of chaotic dynamics. Our results unify subsystem ETH, weak ETH, and macroscopic observables under a single framework, offering new insights into thermalization mechanisms.
\end{abstract}

%\pacs{03.65.Yz}
%\keywords{Lieb-Robinson bound}

\maketitle

\section{Introduction}\label{INTRO}
The equilibration and thermalization of isolated quantum systems are central to understanding how quantum statistical mechanics emerges from unitary quantum dynamics. The eigenstate thermalization hypothesis (ETH) \cite{Rigol16, Deu18} proposes that expectation values of quantum observables in energy eigenstates should approximate thermal values, providing a potential thermalization mechanism. Deutsch's foundational work \cite{Deu91} introduced random perturbations to closed quantum systems, suggesting ETH validity under chaotic dynamics. Srednicki's later formulation using Berry's conjecture \cite{ Sre94, Sre99} also assumes that the eigenstates of the system (i.e., the wavefunctions) are superposition of random plane waves. While these randomness-based frameworks successfully derive thermalization, they risk obscuring the underlying physical origins of thermalization phenomena.

Subsequent ETH studies \cite{BKL10, Mori, LY15, T16, IKS17, DLL18, KS20, MB21, MVY22} abandoned random Hamiltonian assumptions, instead deriving statistical properties from quantum principles.  These investigations focus on demonstrating ``strong (weak) ETH". The strong (weak) ETH posits that all (or almost all) energy eigenstates exhibit thermal properties with respect to local observables, and it postulates that the local distinguishability between an eigenstate and a universal density matrix decreases exponentially with system size. It is generally believed that strong ETH does not apply to integrable or many-body localized systems. By contrast, the weak ETH emphasizes eigenstate typicality \cite{LY15}, where average distinguishability between an eigenstate and the canonical ensemble decays exponentially or algebraically with system size. Notably, weak ETH can be proven for generic translation-invariant systems, including integrable cases \cite{Mori, IKS17, KS20}.

Mathematical ETH proofs \cite{E1, IKS17, KS20} typically define a distinguishability measure based on an arbitrary observable, and then demonstrate that the distinguishability between the eigenstate and the canonical ensemble approaches zero in the thermodynamic limit, regardless of the observable's specific form. This motivates our development of an observable-independent measure. While trace norms \cite{DLL18} offer one such measure, their analytical complexity hinders deeper analysis. Our prior work \cite{HG24} explored ETH using Belavkin-Staszewski relative entropy and the variance under a special local operator. However, we did not fully appreciate the significance of certain formulas, nor conduct numerical calculations on specific models, resulting in weak ETH conclusions with algebraic decay. 

Here we significantly advance this approach by establishing fundamental properties of our variance measure and performing preliminary numerical verification. We begin in Section. \ref{II} by bridging the variance measure to conventional ETH frameworks, demonstrating its superiority in unifying subsystem ETH, weak ETH, and macroscopic observables. In Section. \ref{TODOD}, we derive a universal trade-off equation that rigorously couples diagonal and off-diagonal elements, providing a model-agnostic foundation for analyzing ETH compliance.  In Section \ref{NumR}, we numerically validate the trade-off relation using a tunable Ising spin chain, where varying the longitudinal field allows controlled switching between integrable and non-integrable regimes. By tracking the ratio between diagonal element and its corresponding average off-diagonal elements, we decode how integrability impedes diagonal suppression while leaving off-diagonal averages universally suppressed. Finally, in Section. \ref{CD}, we conclude the paper and discuss related issues.

%%%\cite{MVY22}考虑了不可积系统的scar states,这些low entanglement态会破坏ETH。但只要加入微弱的相互作用，这些scar states的概率就会双指数减小。
% We begin in Section \ref{II} with preliminary results about ETH, subsystem ETH, weak ETH and ETH with the average observable. We show how the variance relates to commonly used measures and demonstrate how it can be extended to the average observable. In Section \ref{TODOD}, we prove  the existence of a universal trade-off equation for the diagonal and off-diagonal elements of variance. This trade off relationship can be extended to the discussion of eigenstate typicality and the average observable. In Section \ref{NumR}, we examine the trade-off relation and the strong (weak) ETH by numerically simulating an Ising spin chain with a transverse and longitudinal magnetic field. Finally, in Section \ref{CD}, we conclude the paper and discuss related issues.

% This work makes three key advances: (1) introduces an observable-independent ETH measure with superior analytical tractability; (2) establishes rigorous diagonal-off-diagonal trade-off relations; (3) provides numerical evidence differentiating integrable and chaotic thermalization pathways. Our framework offers new insights into thermalization mechanisms while resolving previous measure limitations.

\section{Preliminaries}\label{II}
This section reviews fundamental concepts of ETH and subsystem ETH. We discuss weak ETH with eigenstate typicality as presented in \cite{LY15, T16, IKS17, MVY22}, along with eigenstate thermalization involving macroscopic observables from \cite{IKS17, KS20}.

\subsection{ETH and subsystem ETH}
Consider a closed quantum system $B$ governed by Hamiltonian $H$, where $N$ denotes system size. Let $\{\ket{E_i}\}_{i=1}^{D_N}$ represent the Hamiltonian's eigenstates with corresponding eigenvalues $E_i$, satisfying $H\ket{E_i}=E_i\ket{E_i}$. For a spatial subsystem $B_1$, one can define a local distinguishability measure for any few-body observable $A^{B_1}$ as
\begin{equation}\label{ETH}
    d^{(1)}(\Pi^{ij}_{B},\rho_B;A^{B_1}):= \abs{ \Tr[A^{B_1}(\ket{E_i}\bra{E_j}-\rho_B\delta_{ij})]}
\end{equation}
to quantifies the local (in)distinguishability between the energy eigenstates and the density matrix  $\rho_B$. Following local ETH \cite{Rigol16}, this measure follows
\begin{equation}\label{CMETH}
   d^{(1)}(\Pi^{ij}_{B},\rho_B(E);A^{B_1})= e^{-S(E)/2}\abs{f_A(E,\omega)R_{ij}}, 
\end{equation}
where $\rho_B(E)$ denotes a universal density matrix varying smoothly with mean energy $E:=(E_i+E_j)/2$, $f_A(E,\omega)$ represents a smooth function of $E$ and energy difference $\omega:=E_j-E_i$, and $R_{ij}$ stands for a random variable with zero mean and unit variance.

In the measure \eqref{ETH}, the distinguishability depends on the chosen observable. For an observable-independent measure, a more straightforward approach is to consider the following measure:
\begin{equation}\label{subETH}
    d^{(2)}(\sigma_{B_1}^{ij},\rho_{B_1}):= \norm{\sigma_{B_1}^{ij}-\rho_{B_1}\delta_{ij}},
\end{equation}
where $\sigma_{B_1}^{ij}:=\Tr_{\bar{B}_1}\Pi^{ij}_{B}=\Tr_{\bar{B}_1}\ket{E_i}\bra{E_j}$ denotes the reduced matrix on $B_1$, while $\rho_{B_1}:=\Tr_{\bar{B}_1}\rho_B$ represents the subsystem's reduced density matrix. The trace norm is defined as $\norm{O}:=\frac{1}{2}\Tr\sqrt{O^\dagger O}$. Subsystem ETH postulates $ d^{(2)}(\sigma_{B_1}^{ij},\rho_{B_1}(E))\sim\mathcal{O}(e^{-S(E)/2})$, constituting a stronger condition than local ETH \eqref{CMETH}. This hierarchy is supported by the inequality \cite{DLL18}:
\begin{equation}\label{Rd12}
    d^{(1)}(\Pi^{i}_{B},\rho_B;A_{B_1})\leqslant\sqrt{\norm{\sigma_{B_1}^{i}-\rho_{B_1}}_1\Tr[(\Pi^{i}_{B}+\rho_B)A_{B_1}^2]},
\end{equation}
where $\Pi^i_{B}:=\ket{E_i}\bra{E_i}$, $\sigma^{i}:=\sigma^{ii}$, and Schatten $p$-norm $\norm{T}_p:=\left\{\Tr[(T^\dagger T)^{p/2}]\right\}^{1/p}$. The Schatten $1$-norm (trace norm) relates to trace distance through $\norm{\cdot}_1=2\norm{\cdot}$.

While the measure \eqref{subETH} maintains conceptual simplicity, its practical implementation presents challenges due to analytical intractability and limited comparability with related quantities. The Schatten 2-norm (Frobenius norm) offers distinct advantages over the Schatten 1-norm (trace norm) through enhanced computational tractability stemming from its smoothness and differentiable properties. Substituting the trace distance in \cref{subETH} with this alternative norm yields
\begin{equation}
    d^{\prime(2)}(\sigma_{B_1}^{ij},\rho_{B_1}):= \norm{\sigma_{B_1}^{ij}-\rho_{B_1}\delta_{ij}}_2.
\end{equation}
While computationally favorable, this modified measure produces less stringent conditions and insufficiently constrains $d^{(1)}$ magnitude. We therefore propose an alternative approach employing a more sophisticated measure through rescaling transformations. Introducing the operator rescaling map $\mathcal{J}^{\alpha}_{\rho}(\cdot):=\rho^\alpha (\cdot)\rho^{\alpha}$, we reformulate the components of \eqref{subETH} as
\begin{equation}
    \sigma_{B_1}^{ij}-\rho_{B_1}\delta_{ij}= \mathcal{J}_{\rho_{B_1}}^{1/2} (O^{B_1}_{ij}-\braket{O^{B_1}_{ij}}_{\rho_B}),
\end{equation}
where the transformed local operator becomes $O^{B_1}_{ij}:=\mathcal{J}_{\rho_{B_1}}^{-1/2} (\sigma_{B_1}^{ij})$ with ensemble average $\braket{O^{B_1}_{ij}}_{\rho_B}:=\Tr(O^{B_1}_{ij}\rho_B)=\delta_{ij}$. This rescaling procedure necessitates strict positivity of $\rho_{B_1}$ due to inverse operator involvement – a condition naturally satisfied by canonical and microcanonical ensembles under consideration. Applying H\"{o}lder's inequality establishes the crucial relationship
\begin{align}\label{d2d3r}
    2 d^{(2)}(\sigma_{B_1}^{ij},\rho_{B_1})= \norm{\mathcal{J}_{\rho_{B_1}}^{1/2} (O^{B_1}_{ij}-\braket{O^{B_1}_{ij}}_{\rho_B})}_1\notag\\
     \leq \norm{\rho_{B_1}^{1/2} (O^{B_1}_{ij}-\braket{O^{B_1}_{ij}}_{\rho_B})}_2 \times\norm{\rho_{B_1}^{1/2}}_2\notag\\
     =\sqrt{\Tr \left[(O^{B_1}_{ij}-\braket{O^{B_1}_{ij}}_{\rho_B})^\dagger\rho_B(O^{B_1}_{ij}-\braket{O^{B_1}_{ij}}_{\rho_B})\right]}\notag\\
     = \sqrt{V(\rho_B,O^{B_1}_{ij})}=:d^{(3)}(\sigma_{B_1}^{ij},\rho_{B_1}),
 \end{align} 
where the generalized variance measure is defined as:
\begin{equation}\label{VAR}
    V(\rho,A):=\Tr [(A-\braket{A}_{\rho})^\dag\rho(A-\braket{A}_{\rho})]=\norm{\rho^{1/2} (A-\braket{A}_{\rho})}^2_2 .
\end{equation}
The variance expression in $d^{(3)}$ admits an operational reformulation through algebraic manipulation
\begin{align}\label{VCVF}
    V(\rho_B,O^{B_1}_{ij}) =\norm{(\sigma^{ij}_{B_1}-\delta_{ij}\rho_{B_1})\rho^{-1/2}_{B_1}}^2_2\notag\\
   = \Tr(O^{B_1\dag}_{ij}\rho_{B_1}O^{B_1}_{ij})-\delta_{ij}=\Tr(\sigma^{ij}_{B_1}\rho^{-1}_{B_1}\sigma^{ij\dag}_{B_1})-\delta_{ij}.
 \end{align}
A critical relationship emerges between measures  $d^{(3)}$ and measure $d^{(1)}$, paralleling inequality (\ref{Rd12}):
\begin{align}\label{d3d1}
   [ d^{(1)}(\Pi^{ij}_{B},\rho_B;A^{B_1})]^2=\abs{\Tr[(\sigma^{ij}_{B_1}-\delta_{ij}\rho_{B_1})A^{B_1}]}^2\notag\\
   =\abs{\Tr[(\sigma^{ij}_{B_1}-\delta_{ij}\rho_{B_1})(A^{B_1}-\braket{A^{B_1}}_{\rho_B})]}^2\notag\\
   \leq \norm{(\sigma^{ij}_{B_1}-\delta_{ij}\rho_{B_1})\rho^{-1/2}_{B_1}\rho^{1/2}_{B_1}(A^{B_1}-\braket{A^{B_1}}_{\rho_B})}_1^2 \notag\\
   \leq \norm{(\sigma^{ij}_{B_1}-\delta_{ij}\rho_{B_1})\rho^{-1/2}_{B_1}}^2_2\times \norm{\rho^{1/2}_{B_1}(A^{B_1}-\braket{A^{B_1}}_{\rho_B})}^2_2\notag\\
   =[d^{(3)}(\sigma_{B_1}^{ij},\rho_{B_1})]^2\times V(\rho_B,A^{B_1}).
\end{align}
In this inequality, the upper bound of $d^{(1)}$ is determined by two variances. One is state-dependent variance $V(\rho_B,O^{B_1}_{ij})$, which is independent of the specific observable $A$ and determined by the intrinsic properties of the energy eigenstate. The other is the observable-dependent variance $V(\rho_B,A^{B_1})$, which is independent of eigenstate and determined by the specific observable $A$. Although $d^{(3)}$ constitutes a stronger measure, the derived inequality \cref{d3d1} can yield tighter bounds under certain circumstances (a particularly common scenario when considering the ETH). This can be demonstrated through H\"{o}lder's inequality:
\begin{equation}
    \norm{(\sigma^{i}_{B_1}-\rho_{B_1})\rho^{-1/2}_{B_1}}^2_2 \leq  \norm{\sigma^{i}_{B_1}-\rho_{B_1}}^2_1 \times \norm{\rho^{-1/2}_{B_1}}^2_\infty.
\end{equation}
By comparing \cref{Rd12} with \cref{d3d1}, we observe that the upper bound provided by \cref{d3d1} becomes tighter whenever $\norm{\sigma^{i}_{B_1}-\rho_{B_1}}_1 \times \norm{\rho^{-1/2}_{B_1}}^2_\infty \leq 1$. This condition is naturally satisfied in ETH-considered systems: the trace norm $\norm{\sigma^{i}_{B_1}-\rho_{B_1}}_1$ typically decays exponentially with system size, while the spectral norm $\norm{\rho^{-1/2}_{B_1}}_\infty$ exhibits minimal scaling dependence. Numerical computations for specific models (as shown in \cref{FIG6}) further validate this analysis. The plot demonstrates that for all energy eigenstates, the variance satisfies $V(\rho_B,O^{B_1}_{ii}) \sim \norm{\sigma^{i}_{B_1}-\rho_{B_1}}^2_1 < 1$. This empirical evidence confirms that \cref{d3d1} indeed provides a tighter bound compared to \cref{Rd12} in practical implementations.

% 在常见的ETH形式\cref{CMETH}中,一个非常重要的限制就是$R_{ij}$的方差是$1$. 这一性质是由对于大量的态指标$\{i,j\}$统计平均得到的。根据\cref{CMETH,d3d1},应当有
% \begin{align}
%     \mathbb{E} [ d^{(1)}(\Pi^{ij}_{B},\rho_B(E);A^{B_1})]^2=e^{-S(E)}\abs{f_A(E,\omega)}^2\notag\\
%     \leq  V(\rho_B(E),A^{B_1})\times  \mathbb{E}  [V(\rho_B(E),O^{B_1}_{ij})].
% \end{align}
% 测度$d^{(1)}$中$R_{ij}$的方差对应到测度$d^{(3)}$中，则表现为$V(\rho_B(E),O^{B_1}_{ij})$的平均。而$d^{(1)}$中光滑函数$f_A(E,\omega)$的性质应当能被$V(\rho_B(E),O^{B_1}_{ij})$统计得到的光滑函数所体现。对应的，也可以通过分析$V(\rho_B(E),O^{B_1}_{ij})$来得到光滑函数$f_A(E,\omega)$的性质。$R_{ij}$的方差和光滑函数$f_A(E,\omega)$的性质可用于two-point correlator的研究中\cite{FK19,MS19}。

The standard ETH framework in \cref{CMETH} imposes a critical constraint through the unit variance requirement of random variables $R_{ij}$, established via statistical averaging across an extensive eigenstate ensemble $\{i,j\}$. Combining \cref{CMETH} with \cref{d3d1} yields the inequality
\begin{align}
    \mathbb{E} \left[ d^{(1)}(\Pi^{ij}_{B},\rho_B(E);A^{B_1}) \right]^2 = e^{-S(E)}\abs{f_A(E,\omega)}^2 \notag\\
    \leq   \mathbb{E}  \left[V(\rho_B(E),O^{B_1}_{ij})\right]\times V(\rho_B(E),A^{B_1}) .
\end{align}
According to this relationship, the variance characterization of $R_{ij}$ in the $d^{(1)}$ measure manifests as the ensemble average of $V(\rho_B(E),O^{B_1}_{ij})$ in the $d^{(3)}$ measure. Notably, the smooth functional behavior of $f_A(E,\omega)$ within the $d^{(1)}$ framework should be statistically recoverable from the $V(\rho_B(E),O^{B_1}_{ij})$ measurements. This correspondence implies that the characteristic features of the smooth function $f_A(E,\omega)$ can alternatively be extracted through systematic analysis of $V(\rho_B(E),O^{B_1}_{ij})$. Both the variance properties of $R_{ij}$ and the functional characteristics of $f_A(E,\omega)$ find significant applications in the investigation of two-point correlation functions \cite{FK19,MS19}. Research into these aspects presents a potential application of the properties of $V(\rho_B(E),O^{B_1}_{ij})$.

This work employs $d^{(3)}$ as the principal measure for ETH characterization, guided by its unique advantages: 1. Hierarchical Superiority: As rigorously established in \cref{d2d3r}, $d^{(3)}$ provides a theoretical upper bound for preceding measures ($d^{(1)}, d^{(2)}, d^{\prime(2)}$), establishing it as the most stringent ETH criterion within this framework; 2. Operational Versatility: Operating within a mathematically tractable variance framework, $d^{(3)}$ enables systematic connections with thermodynamic observables through \cref{d3d1}; 3. Information Completeness: The variance structure $V(\rho_B,O^{B_1}_{ij})$ intrinsically encodes both diagonal and off-diagonal correlations, providing unified access to ETH signatures across energy eigenstate ensembles.  This tripartite rationale - bounding capability, analytical flexibility, and informational completeness - positions $d^{(3)}$ as our central investigative tool for subsequent thermalization analysis.  

% In this paper, we will focus on the measure $d^{(3)}$. As indicated by \cref{d2d3r}, $d^{(3)}$ serves as an upper bound for the other measures mentioned previously, making it the strongest ETH measure among them. Additionally, because variance is easily transformed, it facilitates connections with other quantities, which will be demonstrated later.

% \begin{table*}[!ht]
%     \centering
%     \begin{tabular}{|l|l|l|l|l|}
%     \hline
%         ETH measure & $d^{(1)}$ & $d^{(2)}$ &$d^{\prime(2)}$ & $d^{(3)}$ \\ \hline
%         Is it observable-independent?  & No & Yes & Yes & Yes \\ \hline
%         Are off-diagonal elements related to $\rho_B$? & No & No & No & Yes \\ \hline
%         Trade-off relation & \cref{d1trade} & No equation & \cref{d2ptrade} & \cref{DGODGRM}  \\ \hline
%         The strength of measure & Weak & Medium &  Weak & Strong\\ \hline
%         Is the trade-off state-independent? & No & No & No & Yes \\ \hline
%     \end{tabular}
%     \caption{Characteristics of different ETH measures. The strength of the measure here means that stricter criteria can establish upper bounds for weaker measures, thereby supporting more robust formulations of the ETH.
%     }
% \end{table*}

\begin{table*}[!ht]
    \centering
    \caption{Comparative Analysis of ETH Measurement Criteria}
    \label{tab:ETH_measures}
    \begin{tabular}{@{}lcccc@{}}
        \toprule
        Property & $d^{(1)}$ & $d^{(2)}$ & $d^{\prime(2)}$ & $d^{(3)}$ \\ 
        \midrule
        Observable independence & \ctimes & \cmark & \cmark & \cmark \\
        $\rho_B$-dependent off-diagonals & \ctimes & \ctimes & \ctimes & \cmark \\
        Trade-off formulation & \cref{d1trade} & N/A & \cref{d2ptrade} & \cref{DGODGRM} \\
        Measurement hierarchy & Weak & Intermediate & Weak & Strong \\
        State-independent trade-off & \ctimes & \ctimes & \ctimes & \cmark \\
        \bottomrule
    \end{tabular}\\
    \vspace{0.2cm}
    \footnotesize{
    \textbf{Legend:} 
    \cmark~= Property satisfied, 
    \ctimes~= Property not satisfied, 
    N/A = Not applicable. \\
    Measurement hierarchy reflects the relative strength in establishing ETH bounds, where stronger measures impose stricter constraints.
    }
\end{table*}

\subsection{Effect of eigenbasis choice in degenerate systems}\label{EECDS}
Consider \( k \)-fold degenerate eigenstates \(\{\ket{\psi_1}, \dots, \ket{\psi_k}\}\) with energy \( E \). Under a unitary transformation \( U \), any superposition state \(\ket{\psi'} = U\ket{\psi}\) remains an eigenstate with the same energy. However, the measure \( d(\sigma_{B_1}^{ij},\rho_{B_1}) \) can change under the new eigenbasis $\{\ket{\psi'}\}$. Here, we analyze how different choices of eigenbasis affect the measure \( d^{(3)} \).

First, note that different eigenbases yield equivalent contributions to the reduced ensemble:
\begin{align}
\sum_{i=1}^{k} \Tr_{\bar{B}_1}\ket{\psi_i}\bra{\psi_i} = \Tr_{\bar{B}_1}\left(\sum_{i=1}^{k}\ket{\psi_i}\bra{\psi_i}\right)\notag\\
 = \Tr_{\bar{B}_1}\left(\sum_{i=1}^{k}\ket{\psi'_i}\bra{\psi'_i}\right) = \sum_{i=1}^{k} \Tr_{\bar{B}_1}\ket{\psi'_i}\bra{\psi'_i}. 
\end{align}
Consequently, the \( \rho_{B_1} \) in measures \( d^{(3)} \) remains invariant under superposition choices. As for the elements \( \sigma_{B_1}^{ij} = \Tr_{\bar{B}_1}\ket{\psi_i}\bra{\psi_j} \), when considering different bases \(\{\ket{\psi'}\}\). The corresponding measure satisfies:
\begin{align} \label{SPINFLU}
V(\rho_B, O^{\prime B_1}_{ij}) &= \norm{\left(\sigma^{\prime ij}_{B_1} - \delta_{ij}\rho_{B_1}\right)\rho^{-1/2}_{B_1}}^2_2 \notag\\
&= \norm{\left(\sum_{ml} U_{im}U^\dagger_{lj}\sigma^{ml}_{B_1} - \delta_{ij}\rho_{B_1}\right)\rho^{-1/2}_{B_1}}^2_2 \notag \\
&\leq \delta_{ij} \sum_{m} |U_{im}|^2 \norm{\left(\sigma^{mm}_{B_1} - \rho_{B_1}\right)\rho^{-1/2}_{B_1}}^2_2 \notag \\
&\quad + \sum_{m \neq l} |U_{im}U^\dagger_{lj}| \cdot \norm{\sigma^{ml}_{B_1}\rho^{-1/2}_{B_1}}^2_2\notag\\
&\leq \delta_{ij}\max_m V(\rho_B, O^{B_1}_{mm})+k \max_{m\neq l} V(\rho_B, O^{B_1}_{ml}),
\end{align}
where in the last inequality we have used the property of unitary transformations and
\begin{align}
    \sum_{m \neq l} |U_{im}U^\dagger_{lj}|\leq  (\sum_{m } |U_{im}|)(\sum_{l }|U_{jl}|)\notag\\
    \leq (\sum_{m } 1 )(\sum_{l }|U_{jl}|^2)=k.
\end{align}
As long as the degeneracy count \( k \) grows significantly slower than the exponential decay constant of   \( d^{(3)} \), the superposition-induced discrepancies in the measures become negligible. Similar bounds apply to measures \( d^{(2)} \), \( d^{\prime(2)} \), and \( d^{(1)} \), which can be analyzed using a method similar to \cref{SPINFLU}. Notably, \( d^{(1)} \), being observable-dependent, may exhibit distinct behavior since degenerate states may remain indistinguishable for specific observables. 

\subsection{Weak ETH with eigenstate typicality}\label{WETHwET}

While theoretically robust, the strong ETH formulation may impose excessive constraints in practical scenarios. This limitation becomes particularly evident when examining the standard expression \eqref{CMETH}, which regulates only the variance of $R_{ij}$ through ensemble averaging, leaving maximal fluctuation magnitudes unbounded. Direct characterization of $R_{ij}$'s intrinsic stochastic properties presents significant analytical hurdles due to its implicit state dependence. A more operationally viable approach emerges through statistical ensemble analysis of observable concentration.

Consider an energy eigenstate ensemble for composite system $B$ described by density matrix $\rho_B = \sum_j p_j \Pi^j_{B}$. The ``weak'' ETH framework transitions focus to establishing statistical regularity and quantifying observable typicality through probabilistic measures. This probabilistic perspective employs the concentration measure \cite{HG24}:
% The strong ETH is in some ways too strong. In the commonly used ETH form (\ref{CMETH}), there is no limit on the maximum value of $\abs{R_{ij}}$, only its variance. However, the randomness of $R$ in (\ref{CMETH}) is not easy to analyze. A suitable approach is to study its statistical behavior.  Consider an ensemble consisting of energy eigenstates of the total system $B$, denoted by $\rho_B=\sum_j p_j \Pi^j_{B}$.  In many investigations of the ``weak'' ETH, the goal is to derive statistical properties and quantify probabilistic typicality or concentration with respect to the measure $d^{(1)}$. For this, the following expression is used \cite{HG24}
\begin{equation}\label{IDTYP}
    \Delta(\rho_B,A^{B_1}):=\int   [d^{(1)}(\Pi^{i}_{B},\rho_B;A^{B_1})]^2 p_d,
\end{equation}
where the probability distribution $p_d$ is derived from the  $p_j$'s through a Radon-Nikodym derivative. According to the Chebyshev inequality, we have
\begin{equation}\label{CHEBY}
    P_{\rho}(d^{(1)}(\Pi^{i}_{B},\rho_B;A^{B_1})\geqslant \epsilon)\leqslant\frac{\Delta(\rho,A)^2}{\epsilon^2},\quad\forall\epsilon\in\mathbb{R^+}.
\end{equation}
Thus, when $ \Delta(\rho,A)$ is very small,  the expectation of the projectively  measured observable would concentrate on the expectation of observable calculated with respect to the state $\rho_B$. For any microcanonical ensemble  $\rho^\text{mc}$, if 
\begin{equation}\label{14}
\Delta(\rho^\text{mc},A^{B_1})\sim\mathcal{O}(N^{-\alpha}),
\end{equation}
with $0<\alpha<1$, which means that the expectations of a local observable $A^{B_1}$ with respect to the results of local measurements concentrate around the expectation of $A^{B_1}$ with respect to $\rho^\text{mc}$, then it follows that each pure state $\Pi^{i}_{B}$ cannot be distinguished from the microcanonical state  $\rho^\text{mc}$ in the large $N$ limit. This result characterizes the {\it weak ETH with eigenstate typicality} \cite{IKS17}. By substituting the measure  $d^{(1)}$ with other measures, different forms of the weak ETH can be obtained.

\subsection{The average observable}

Subsystems in many-body quantum systems inherently exhibit state fluctuations. Even when governed by translationally invariant Hamiltonians, only equilibrium ensembles like the Gibbs state (canonical ensemble) and microcanonical ensemble preserve strict translational symmetry. Individual energy eigenstates, however, manifest residual spatial fluctuations that enhance their distinguishability compared to translation-invariant ensembles.

To systematically address these fluctuations, we partition system $B$'s lattice into $\mathcal{C} = |B|/|B_1|$ identical subsystems $B_k$ ($k=1,...,\mathcal{C}$), where $|B|$ denotes total lattice sites. In order to eliminate the influence of
spatial fluctuations, or consider the macroscopic observable \cite{KS20}, one can use the spatially average observable  
\begin{equation}\label{TFLA}
    A^{B}:=\frac{1}{\mathcal{C}}\sum_i A^{B_i},
\end{equation}
where $A^{B_i}$ represents the translational copy of $A^{B_1}$. This regularization procedure effectively projects local operators onto the translation-symmetric subspace, reconciling eigenstate fluctuations with thermodynamic predictability. Substituting $A^{B}$ into \cref{ETH} gives
\begin{align}\label{SDTI}
    d^{(1)}(\Pi^{ij}_{B},\rho;A^B)=\abs{\frac{1}{\mathcal{C}}\sum_k\Tr [(\sigma^{ij}_{B_k} -\delta_{ij}\rho_{B_k}) A^{B_k}]} \notag\\
    =\abs{\Tr [(\frac{1}{\mathcal{C}}\sum_k\sigma^{ij,k}_{B_1} -\delta_{ij}\overline{\rho}_{B_1}) A^{B_1}]}\notag\\
    =d^{(1)}(\frac{1}{\mathcal{C}}\sum_k\sigma^{ij,k}_{B_1},\overline{\rho}_{B_1};A^{B_1}),
\end{align}
where  $\sigma^{ij,k}_{B_1}$ is the translational copy of $\sigma^{ij}_{B_k}$ and $\overline{\rho}_{B_1}:=\Tr_{\bar{B}_1}\overline{\rho}_B$. The density matrix $\overline{\rho}_B:=(\sum_k \hat{T}_k \rho_{B})/\mathcal{C} $ and $\hat{T}_k$ is the translation operator that maps $B_k$ to $B_1$.  According to \cref{SDTI,d3d1}, an upper bound can be given by 
\begin{align}\label{avgoVAR}
   [ d^{(1)}(\Pi^{ij}_{B},\rho;A^B)]^2\notag\\
    \leq \norm{(\frac{1}{\mathcal{C}}\sum_k\sigma^{ij,k}_{B_1} -\delta_{ij}\overline{\rho}_{B_1})\overline{\rho}^{-1/2}_{B_1}\overline{\rho}^{1/2}_{B_1}(A^{B_1}-\braket{A^{B_1}}_{\overline{\rho}_B})}_1^2 \notag\\
    \leqslant V(\overline{\rho}_B,\overline{O}^{B_1}_{ij})\times V(\overline{\rho}_B,A^{B_1}),
\end{align}
where $\overline{O}^{B_1}_{ij}:=\frac{1}{\mathcal{C}}\sum_{k}\mathcal{J}_{\overline{\rho}_{B_1}}^{-1/2} (\sigma^{ij,k}_{B_1})$, and $\sigma^{ij,k}_{B_1}$ are the translational copies of $\sigma^{ij}_{B_k}$.

\section{Trade-off between diagonal and off-diagonal elements}\label{TODOD}
In this section, we examine the relationship between the diagonal elements \( d(\sigma_{B_1}^{ii}, \rho_{B_1}) \) and the off-diagonal elements \( d(\sigma_{B_1}^{ij; i \neq j}, \rho_{B_1}) \) in the context of the ETH metric. For clarity, we emphasize that throughout this work, the terms ``diagonal elements" and ``off-diagonal elements" specifically refer to these defined components of the ETH metric characterization, rather than denoting direct matrix elements of physical observables.

\subsection{Trade-off between diagonal and off-diagonal elements}\label{LOCase}
For any ensemble $\rho_B$ with strictly positive reduced matrix $\rho_{B_1}=\Tr_{\bar{B}_1}\rho_B$, the diagonal element $V_\text{dg}^{i}:=V(\rho_B,O^{B_1}_{ii})$ and the off-diagonal element $V_\text{off}^{ij}:=V(\rho_B,O^{B_1}_{ij;i\neq j})$ in $d^{(3)}$ are not independent. According to \cref{DGODGR}, they are related by the following equation
\begin{equation}\label{DGODGRM}
    V_\text{dg}^{i}+ \sum_{j\neq i} V_\text{off}^{ij}=\Tr(\rho_{B_1}^{-1})-1,
\end{equation}
where the index \( j \) runs over all states in the Hilbert space excluding \( i \). The summation contains \( D_N - 1 \) terms, whose number scales exponentially with the system size \( N \). However, the right-hand side of \eqref{DGODGRM} depends solely on the local state $\rho_{B_1}$, and it generally grows exponentially with the size of subsystem $N_{B_1}$, while remaining relatively constant with respect to $N$.  This causes the off-diagonal elements to be exponentially suppressed. Considering the average of these off-diagonal elements
\begin{equation}\label{ODA}
    V_\text{off}^{i}:= \frac{1}{D_N-1}\sum_{j\neq i} V_\text{off}^{ij},
\end{equation}
which will decrease exponentially with $N$. However, this exponential decrease in the average of the off-diagonal elements does not imply that the maximum values of the off-diagonal elements and the diagonal element will also decrease exponentially. Although the diagonal elements are not directly suppressed, \cref{DGODGRM} provides a trade-off relation between the diagonal element and the off-diagonal elements. With $ V_\text{dg}^{i}/V_\text{off}^{i}=:f^i(N)$, we can rewrite \cref{ODA} as
\begin{equation}\label{DoDratio}
    V_\text{dg}^{i}=\frac{f^i(N)}{f^i(N)+D_N-1}(\Tr(\rho_{B_1}^{-1})-1).
\end{equation}
On one hand, if the ratio between diagonal and average off-diagonal elements $f^i(N)$ does not increase exponentially, or grows exponentially at a rate much slower than $D_N$, then in the large $N$ limit, the diagonal elements will also be exponentially small. On the other hand, if the diagonal element is not decreasing exponentially, $f^i(N)$ must be growing exponentially. Thus, the exponential growth of $f^i(N)$ is a necessary condition for the diagonal element to avoid exponential decay.

In this paper, we do not delve further into how interactions affect the growth of $f^i(N)$ with $N$. However, as will be evident from the subsequent numerical calculations, the influence of integrability on this growth is crucial. To fully address the underlying theoretical mechanisms and other factors, such as the choice of $\rho_{B_1}$, further investigation is needed. One potential approach involves using H\"{o}lder's inequality
\begin{align}
    V_\text{dg}^{i}=\norm{(\sigma^{i}_{B_1}-\rho_{B_1})\rho^{-1/2}_{B_1}}^2_2\notag\\
    \leq \norm{(\sigma^{i}_{B_1}-\rho_{B_1})(\sigma^{ij}_{B_1})^{-1}}^2_\infty  \norm{\sigma^{ij}_{B_1}\rho^{-1/2}_{B_1}}^2_2.
\end{align}
Combining this result with \cref{ODA}, it can be easily shown that
\begin{equation}\label{fiIEQ}
    f^i(N)\leq \frac{D_N-1}{\sum_{j\neq i} \norm{(\sigma^{i}_{B_1}-\rho_{B_1})(\sigma^{ij}_{B_1})^{-1}}^{-2}_\infty},
\end{equation}
where $\sigma_{B_1}^{ij}=\text{Tr}_{\bar{B}_1}\ket{E_i}\bra{E_j}$.
The right-hand side of inequality (\ref{fiIEQ}) may provide a more straightforward framework to discuss the effects of interactions and the choice of the $\rho_{B_1}$ on $f^i(N)$.

\begin{figure*}
    \centering
    \subfigure[]{\includegraphics[width=0.48\textwidth]{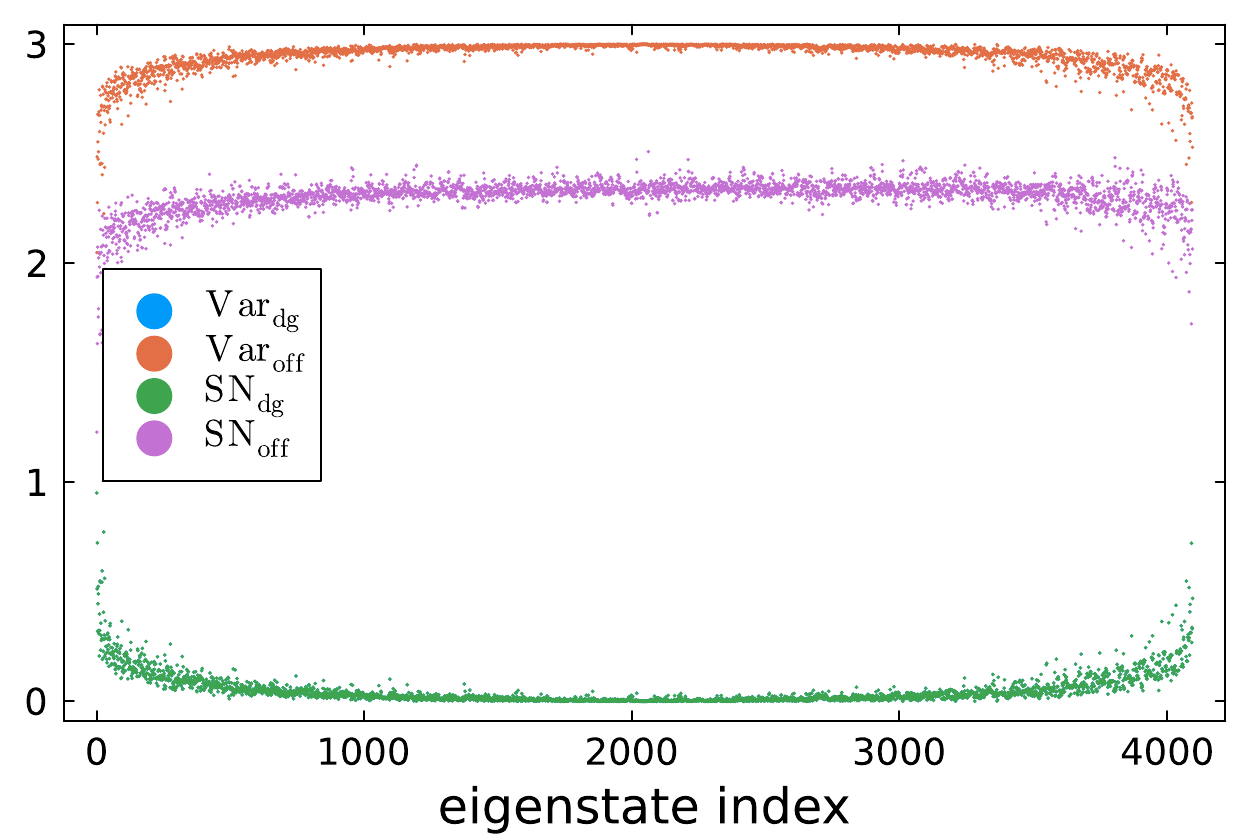}}
    \subfigure[]{\includegraphics[width=0.48\textwidth]{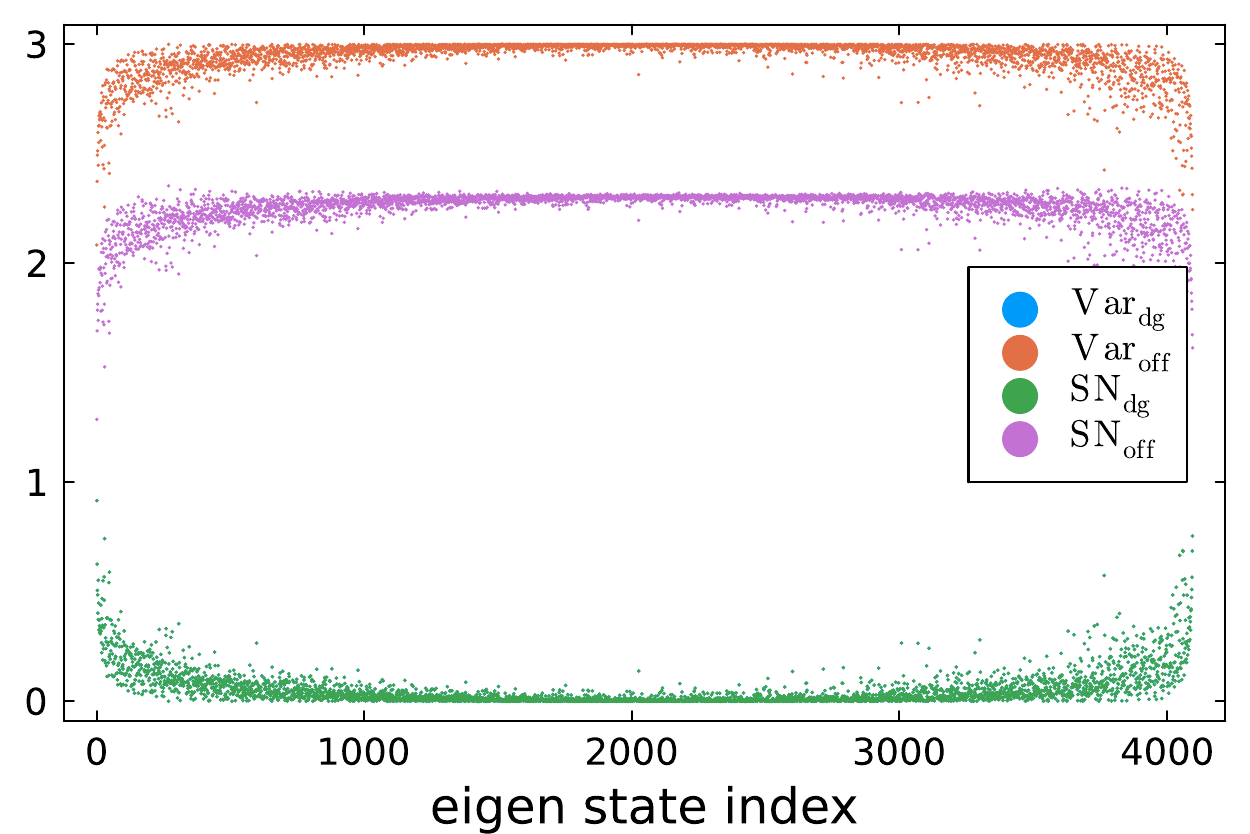}}
    \caption{Diagonal and off-diagonal elements versus eigenstate index. (a) Spin chain under PBC. (b) Spin chain under OBC. In the plot, $\text{Var}_\text{dg}:= V_\text{dg}^{i}$, $\text{Var}_\text{off}:= \sum_{j\neq i} V_\text{off}^{ij}$, $\text{SN}_\text{dg}:=  \norm{\sigma_{B_1}^{i}-\rho_{B_1}}_1^2$ and  $\text{SN}_\text{off}:=  \sum_{j\neq i} \norm{\sigma_{B_1}^{ij}}_1^2$. The variance measures (Var) systematically produces larger values than the trace-norm measures (SN), validating \cref{d2d3r} and the hierarchy in \cref{tab:ETH_measures}. The measure $\text{Var}_\text{dg}$ closely aligns with $\text{SN}_\text{dg}$, making the data points for $\text{Var}_\text{dg}$ overlap with those for $\text{SN}_\text{dg}$. Eigenstates are sorted by energy (low to high). Parameters: $g =1.05$, $h=0.1$, $N = 12$, $N_{B_1}$=1, $\rho_B=\sum_i \Pi^i_B/D_N$.}\label{FIG6}
\end{figure*}

The index $i$ in \cref{DGODGRM} can be chosen arbitrarily, but the right side of the formula is independent of eigenstate index $i$. When the measure $V_\text{dg}$ of a given eigenstate is large, the average measure of its off-diagonal elements $V_\text{off}^{i}$ must be small, and vice versa. Fig. \ref{FIG6} illustrates this trade-off relation, where the energy eigenstates are numerically calculated based on a specific non-integrable system. We will introduce the model used in detail in \cref{NumR}. Since the measure $d^{(2)}$ and $d^{(3)}$ are related by \cref{d2d3r}, the trade-off between the diagonal and off-diagonal elements is also reflected in $d^{(2)}$, although it is not an exact equation relationship like \cref{DGODGRM}.  From the \cref{FIG6}, we observe that the diagonal element is generally much smaller than the sum of the corresponding off-diagonal elements. But this does not mean that $f^i$ is generally much smaller than $1$. On the contrary, since there are $D_N-1$ off-diagonal elements for each diagonal element, $f^i$ is typically much larger than $1$, which will be shown in \cref{FIGsfi}.

The formula \eqref{DGODGRM} holds for any $\rho_B$ as long as $\rho_{B_1}$ is strictly positive. This includes not only the canonical ensemble $\rho^{\text{c}}(\beta):=e^{-\beta H}/Z(\beta)$ and microcanonical ensembles, but also pure states such as $\Pi^i_B$. In the case of pure states, the diagonal element in \cref{DGODGRM} will be zero, leading to
\begin{equation}
   \sum_{j\neq i} V(\Pi^i_B,O^{B_1}_{ij})=\Tr(\sigma_{B_1}^{-1})-1,
\end{equation}
where the sum of all corresponding off-diagonal elements is directly determined by the local state $\sigma_{B_1}=\Tr_{\bar{B}_1}\Pi^i_B$.

The trade-off relationship is not exclusive to the measure $d^{(3)}$; Other measures can also yield similar relationships. For instance, for measure  $d^{(1)}$, it can be shown that
\begin{align}\label{d1trade}
    \sum_j  [d^{(1)}(\Pi^{ij}_{B},\rho_B;A^{B_1})]^2=\sum_j \abs{\bra{E_j}A^{B_1}-\braket{A^{B_1}}\ket{E_i}}^2\notag\\
    =\Tr[\Pi^i_B (A^{B_1}-\braket{A^{B_1}})(A^{B_1}-\braket{A^{B_1}})^\dag].
\end{align}
Compared with \cref{DGODGRM}, the right-hand side of \cref{d1trade} depends on the state index $i$. However, since $(A^{B_1}-\braket{A^{B_1}})(A^{B_1}-\braket{A^{B_1}})^\dag$ is a local operator on $B_1$, when ETH holds,  the right-hand side of the equation will approximate the variance $V(\rho_B,A^{B_1})$ according to \cref{CMETH,VAR}. 

Another key difference lies in whether the measure is observable-independent. The right-hand side of \cref{d1trade} explicitly depends on the observable $A^{B_1}$, contrasting with the observable-independent nature of the measure in \cref{DGODGRM}. For the observable-independent measure $d^{(2)}$, its form is not easily transformed, and thus we have not yet identified a direct trade-off relationship. However, as shown in the numerical calculations for a specific model in \cref{FIG6}, a similar relationship does emerge, though, like in \cref{d1trade}, it is not independent of the state index $i$. For the measure $d^{\prime(2)}$, the following trade-off holds:
\begin{align}\label{d2ptrade}
    \sum_j  [d^{\prime(2)}(\sigma_{B_1}^{ij},\rho_{B_1})]^2\notag\\
    = \sum_j \Tr(\sigma^{ij}_{B_1}\sigma^{ij\dag}_{B_1})+\Tr(\rho_{B_1}^2)-2\Tr(\rho_{B_1}\sigma_{B_1}^{i})\notag\\
    =D_{B_1}+\Tr(\rho_{B_1}^2)-2\Tr(\rho_{B_1}\sigma_{B_1}^{i}),
\end{align}
where the final equation follows from substituting $\rho_{B_1}^{-1}=I_{B_1}$ in \cref{DGODGR}. Here, $D_{B_1}$ denotes the Hilbert space dimension of subsystem $B_1$. The right-hand side of \cref{d2ptrade} also depends on the state index  $i$.

% For a specific energy shell, such as
% \begin{equation}
%     \mathcal{M}_{E,\Delta}=(E-\Delta E/2,E+\Delta E/2),
% \end{equation}
% 笼统来说，会用到下面形式
% \begin{align}
%    \sum_{j\in\mathcal{M}_{E,\Delta}}   [d^{(1)}(\Pi^{ij}_{B},\rho_B(E);A^{B_1})]^2 \notag \\
%     \to \int_{-\Delta E/2}^{\Delta E/2} d\omega e^{-S(E)+S(E+\omega/2)}\abs{f_A(E,\omega)}^2
% \end{align}
% 根据\cref{d3d1},有
% \begin{align}\label{D3asFUP}
%     \sum_{j\in\mathcal{M}_{E,\Delta}}   [d^{(1)}(\Pi^{ij}_{B},\rho_B(E);A^{B_1})]^2 \notag \\
%     \leq  V(\rho_B(E),A^{B_1})\times  \sum_{j\in\mathcal{M}_{E,\Delta}}  V(\rho_B(E),O^{B_1}_{ij})
% \end{align}
% 根据这一不等式，可以由度量$d^{(3)}$的性质来分析光滑函数$f_A(E,\omega)$的上界。此外，根据trade-off关系\cref{DGODGRM},\cref{D3asFUP}中的
% \begin{equation}
%     \sum_{j\in\mathcal{M}_{E,\Delta}}  V(\rho_B(E),O^{B_1}_{ij})=\Tr(\rho_{B_1}^{-1}(E))-1 - \sum_{j\notin \mathcal{M}_{E,\Delta}}V_\text{off}^{ij} .
% \end{equation}
% 这使得我们可以通过能壳外的非对角元来分析光滑函数$f_A(E,\omega)$。

\subsection{Weak ETH}\label{weakETH}
We examine the use of measure $d^{(3)}$ to study the weak ETH with eigenstate typicality by considering both ``diagonal" and ``off-diagonal" probabilistic typicalities. The ``diagonal" probabilistic typicality is given by
\begin{equation}\label{CLQV}
    \braket{V_\text{dg}}:= \sum_{i} p_i  V_\text{dg}^{i}
 \end{equation}
while the ``off-diagonal" probabilistic typicality is
 \begin{equation}\label{CLQVOFF}
     \braket{V_\text{off}}:= \sum_{i} p_i   V_\text{off}^{i}.
  \end{equation}
Given that the relationship in \cref{DGODGRM} holds for each $i$, averaging over the probabilities yields
  \begin{equation}\label{LVVCAN}
    \braket{V_\text{dg}}+ (D_N-1)\times \braket{V_\text{off}}= \Tr(\rho_{B_1}^{-1})-1.
\end{equation}
Similar to the discussion in the previous section, the probability average of off-diagonal elements $ \braket{V_\text{off}}$ should decrease exponentially with $N$. According to \cref{DoDratio}, the  ``diagonal" probabilistic typicality
\begin{equation}
    \braket{V_\text{dg}}=[\Tr(\rho_{B_1}^{-1})-1] \times \sum_{i} p_i \frac{f^i(N)}{f^i(N)+D_N-1},
\end{equation}
and its decay behavior with increasing system size depends on the specific model.

\subsection{The average observable}
If we consider the distinguishability measure $ V(\rho_B,\overline{O}^{B_1}_{ij})$ given by the average observable  $\overline{O}^{B_1}_{ij}=\frac{1}{\mathcal{C}}\sum_{k}\mathcal{J}_{\rho_{B_1}}^{-1/2} (\sigma^{ij,k}_{B_1})$, then according to \cref{avgoDGODGR}, the corresponding diagonal and off-diagonal elements exhibit the following trade-off relationship
  \begin{align}\label{avgotdoff}
    V(\rho_B,\overline{O}^{B_1}_{ii})+\sum_{j\neq i} V(\rho_B,\overline{O}^{B_1}_{ij})=\frac{1}{\mathcal{C}}(\Tr(\rho_{B_1}^{-1})-1) \notag\\
    +\frac{1}{\mathcal{C}^2} \sum_{\beta, \alpha,l,k\neq l}  p'_\alpha \Tr [O^{B_k,\dag}_{\alpha\beta,1}\otimes O^{B_l}_{\alpha\beta,1} (\sigma^{i}_{B_kB_l}-\sigma^{i}_{B_k}\otimes \rho^{1}_{B_l}) ],
\end{align}
where $\rho^{1}_{B_l}$ is the translational copy of $\rho_{B_1}$. Although formula \eqref{avgotdoff} is valid for any $\rho_B$ with strictly positive $\rho_{B_1}$, it connects to \cref{avgoVAR} specifically when using $\overline{\rho}_B$. Thus, all quantities related to $\rho_B$ should be replaced with $\overline{\rho}_B$, which is translation invariant. In such cases  $\overline{\rho}^{1}_{B_l}=\overline{\rho}_{B_l}$. The local term in \cref{avgotdoff} becomes $\Tr(\overline{\rho}_{B_1}^{-1})-1$, which is related to the local state $\overline{\rho}_{B_1}$ and includes an additional factor of $1/\mathcal{C} $. The local term decreases algebraically as $\mathcal{C}$ increases. The second term on the right side of \cref{avgotdoff} is related to the spatial correlation of $\sigma^{i}$ and the difference between $\sigma^{i}_{B_l}$ and $\rho^{1}_{B_l}$ (or $\overline{\rho}_{B_l}$ for translation invariant system).

We can also consider both the average observable and eigenstate typicality. In this case, we need to examine the following ``diagonal'' probabilistic typicality
\begin{equation}\label{CAQV}
    \braket{V_\text{dg}}= \sum_{i} p_iV(\rho_B,\overline{O}^{B_1}_{ii}).
 \end{equation}
And the ``off-diagonal'' probabilistic typicality
 \begin{equation}\label{CAQVOFF}
     \braket{V_\text{off}}= \frac{1}{D_N-1}\sum_{i,j\neq i} p_iV(\rho_B,\overline{O}^{B_1}_{ij}).
  \end{equation}
After averaging the probability of \cref{avgotdoff}, we obtain the following trade-off relationship
  \begin{equation}\label{Gl2LoCo}
    \braket{V_\text{dg}}+(D_N-1)\times\braket{V_\text{off}}=\frac{1}{\mathcal{C}}(\Tr(\rho_{B_1}^{-1})-1)+V^{\text{Avg}}_{\text{Cor}},
\end{equation}
where 
\begin{equation}\label{Vavg}
    V^{\text{Avg}}_{\text{Cor}}:= \frac{1}{\mathcal{C}^2} \sum_{\beta, \alpha,l,k\neq l}  p'_\alpha \Tr [O^{B_k,\dag}_{\alpha\beta}\otimes O^{B_l}_{\alpha\beta} (\rho_{B_kB_l}-\rho_{B_k}\otimes\rho_{B_l}) ],
 \end{equation}
 which is determined by the spatial correlation of $\rho_B$.

 As previously mentioned, when using \cref{avgoVAR}, all quantities in formula \eqref{Gl2LoCo} related to $\rho_B$ must be replaced with $\overline{\rho}_B$, except for $\rho_{B_kB_l}$ and $\rho_{B_k}$ in \cref{Vavg}. If $\rho_B$ is translation invariant, then $\rho_B=\overline{\rho}_B$, and the second term on the right side of \cref{Gl2LoCo} is solely related to the correlation of $\overline{\rho}_B$. If the spatial correlation of $\overline{\rho}_B$ decays exponentially or algebraically fast enough, the right side of \cref{Gl2LoCo} will exhibit algebraic decay of the form $O(N^{-\epsilon})$ \cite{HG24}. Compared to \cref{LVVCAN}, the right side of \cref{Gl2LoCo} depends not only on the local state but also on the rate of correlation decay. Both terms generally decrease algebraically with increasing $N$, consistent with the intuition that spatial averaging smooths and reduces differences (distinguishability). Although this algebraic decay alone does not guarantee strong ETH, it supports the validity of weak ETH with algebraic decay,  even for integrable systems. This algebraic decay aligns with the findings of our previous article \cite{HG24}, though we did not previously account for the $D_N-1$ off-diagonal elements in \cref{Gl2LoCo}, which cause the average off-diagonal elements to be exponentially suppressed and further restrict their values. Consequently, as the right side of \cref{Gl2LoCo} decreases with $N$, the exponential decrease of $\braket{V_\text{off}}$ remains applicable. Additionally, due to the smoothing effect of the average observable, the values of $\braket{V_\text{dg}}$ and $\braket{V_\text{off}}$ here will be smaller than those defined in the \cref{weakETH}.

\section{Numerical Results}\label{NumR}

To validate our theoretical framework, we perform numerical simulations on an Ising spin chain under combined transverse and longitudinal magnetic fields governed by the Hamiltonian:
\begin{equation}\label{HMT}
    H = \sum_{k=1}^{N} \left( -\sigma_z^k \otimes \sigma_z^{k+1} + g \sigma_x^k + h \sigma_z^k \right)
\end{equation}
Notably, this system remains nonintegrable except in the singular cases where either coupling constant $g$ or $h$ vanishes. We perform calculations for both periodic boundary conditions (PBC) and open boundary conditions (OBC). Although PBC are adopted for computational convenience when considering average observable, we emphasize that translation invariance does not constitute an essential requirement for our theoretical framework - most derivations in \cref{II,TODOD} maintain validity even for spatially inhomogeneous systems.

A crucial distinction between the Ising models under PBC and OBC lies in the presence of energy level degeneracy under PBC (absent in OBC for the parameter \( g = 1.05 \) discussed here). While the relationships between different measures in \cref{II} and the trade-off relations in \cref{TODOD} remain unaffected by energy level degeneracy (as evidenced by the theoretical derivation of the trade-off relations and  \cref{FIG6}), degeneracy exerts both direct and indirect influences. Directly, it influences eigenstate thermalization through perturbations that lift the degeneracy. Indirectly, it affects the magnitude of the measures via the superposition of degenerate eigenstates. The extent of this indirect influence can be inferred from the proof in \cref{EECDS}, and thus no further numerical analysis is presented here.

We employ the MPO functions from the ITensor library \cite{FW22} to construct the Hamiltonian matrix. Subsequent exact diagonalization using ITensor's eigen solver yields all energy eigenstates $\{\ket{E_i}\}$ and corresponding eigenvalues $\{E_i\}$. These are utilized to generate both microcanonical and canonical ensembles. After performing partial trace operations to reduce the ensembles, we compute the distinguishability measures $d(\sigma_{B_1}^{ij}, \rho_{B_1})$ for all energy eigenstates. The numerical evaluation of $d^{(3)}$ primarily relies on the formulation provided in \cref{VCVF}. The local distinguishability measures were computed by direct evaluation of reduced density matrices and operator variances without employing matrix product state approximations. 

Adopting parameter values from \cite{DLL18} for direct comparison, we analyze two distinct regimes: (i) $g=1.05$ with $h=0$ (integrable case), and (ii) $g=1.05$ with $h=0.1$ (nonintegrable case). Key thermodynamic quantities including energy level statistics, energy-temperature relationships, and canonical ensemble mutual information are systematically compared in \cref{BPM}. Remarkably, our analysis reveals no significant thermodynamic differentiation between integrable and nonintegrable regimes through these conventional properties.

\begin{figure*}
    \centering
    \subfigure[]{\includegraphics[width=0.48\textwidth]{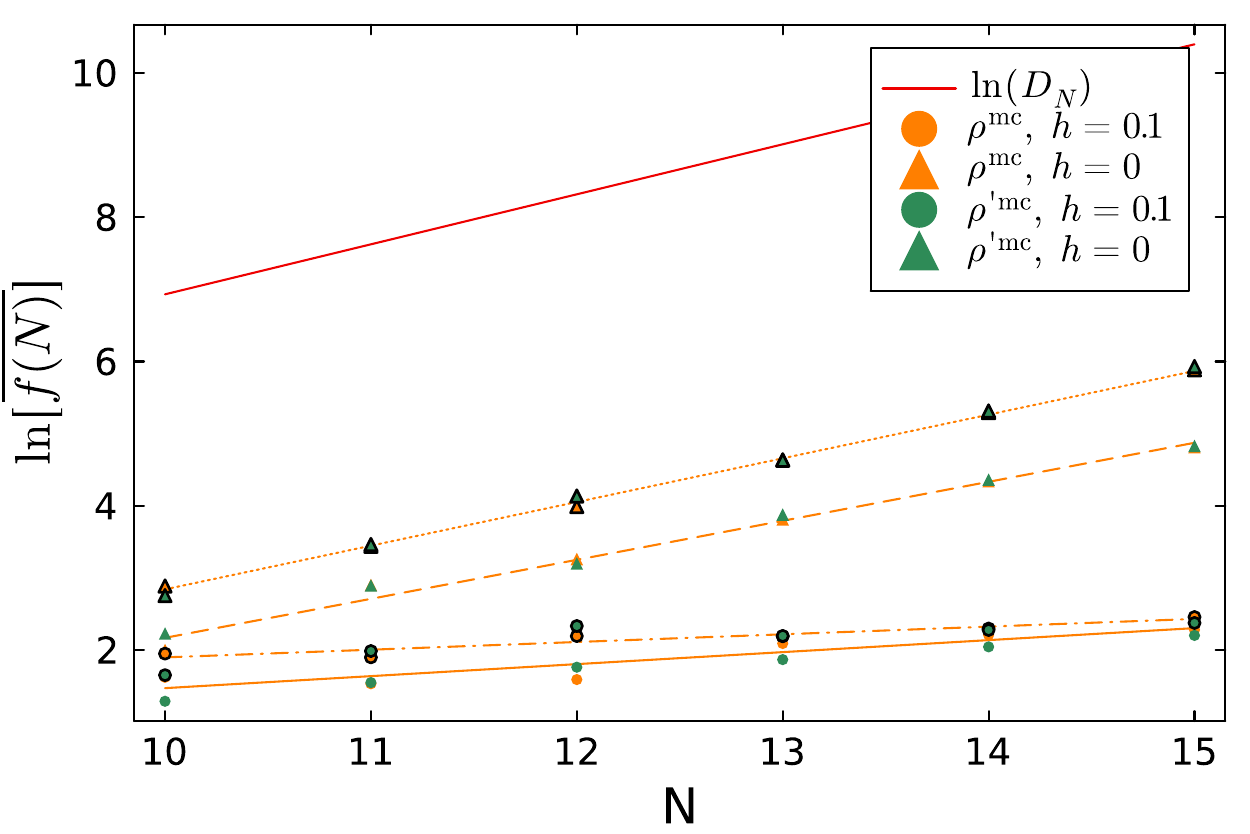}} 
    \subfigure[]{\includegraphics[width=0.48\textwidth]{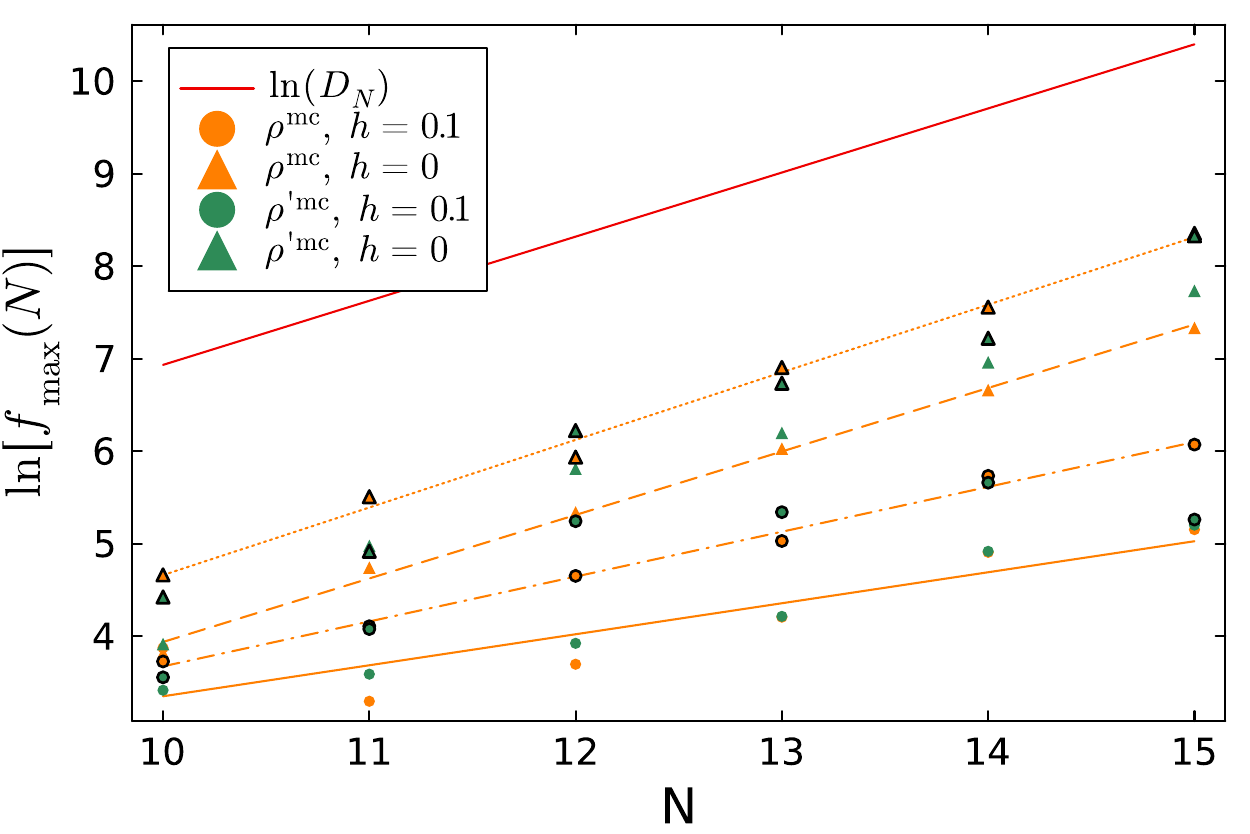}} 
    \caption{Parameters: $g = 1.05$,  $N_{B_1}=1$. In the figure, zero markerstrokewidth corresponds to PBC, whereas non-zero markerstrokewidth indicates OBC. Linear fits are shown exclusively for the energy shell $\mathcal{M}_{E(\beta),\Delta}$ with $\rho_B=\rho^{\text{mc}}$.  (a) The $\overline{f(N)}$ versus $N$. Under PBC, non-integrable systems (solid yellow line) exhibit an exponential growth rate of $0.24\times \ln 2$ (standard error $0.05\times\ln 2$), while integrable systems (dashed yellow line) follow $0.78\times \ln 2$ (standard error $0.05\times\ln 2$). Under OBC, non-integrable systems (dashdot yellow line) exhibit an exponential growth rate of $0.16\times \ln 2$ (standard error $0.03\times\ln 2$), while integrable systems (dashed yellow line) follow $0.88\times \ln 2$ (standard error $0.02\times\ln 2$).  (b) The $f_{\max}(N)$ versus $N$. Under PBC, non-integrable systems show growth at $0.49\times \ln 2$ (standard error $0.13\times\ln 2$), contrasting with integrable systems at $1.00\times \ln 2$ (standard error $0.03\times\ln 2$). Under OBC, non-integrable systems show growth at $0.70\times \ln 2$ (standard error $0.03\times\ln 2$), contrasting with integrable systems at $1.06\times \ln 2$ (standard error $0.04\times\ln 2$).}
    \label{FIGsfi}
\end{figure*}

The integrability of a system profoundly influences ETH-related measures, particularly in governing thermalization dynamics. Through analyzing the measure \([d^{(3)}]^2\) (equivalent to \(V_{\text{(off-)dg}}^{i}\)) for model \cref{HMT} (see \cref{NVETH}), we demonstrate its capability to capture both integrability-dependent suppression of thermalization and system-size-induced exponential decay, consistent with established ETH measures. Under the conventional ETH framework \cref{CMETH}, where ensemble selection depends on energy, we systematically investigated the energy-difference dependence of \(d^{(3)}\) in \cref{IEDTD}. While \(V_{\text{(off-)dg}}^{i}\) retains residual energy dependence due to varying state densities, this effect contributes negligibly to distinguishability compared to ensemble-energy mismatches. 
% Crucially, non-integrable systems exhibit universal exponential suppression of \(V_{\text{(off-)dg}}^{i}\), as evidenced for both local observables (\cref{ETHLM}) and spatially averaged observables (\cref{ETHAO}), satisfying the weak ETH with faster diagonal decay rates than integrable counterparts.  

Due to the significant enhancements in distinguishability resulting from ensemble-energy mismatches, and given that we compute ensembles at fixed energy values, our analysis of diagonal and off-diagonal matrix elements (as discussed here and in \cref{ETHLM,ETHAO}) specifically considers eigenstate indices \(i\) whose corresponding energies lie within the energy shell interval:
\begin{equation}
    \mathcal{M}_{E,\Delta}=(E-\Delta /2,E+\Delta /2),
\end{equation}
where \(\Delta = 0.2N\). Our analysis focuses on energy shells \(\mathcal{M}'_{E=0,\Delta}\) and \(\mathcal{M}_{E(\beta),\Delta}\), with \(E(\beta)\) corresponding to the canonical ensemble \(\rho^\text{c}(\beta = 0.1)\). As demonstrated in \cref{ETHLM,ETHAO}, these shells reveal distinct integrability-dependent signatures of ETH adherence:
\begin{itemize}
    \item \textbf{Non-integrable systems:} These systems exhibit clear adherence to weak ETH. Both the average diagonal elements and off-diagonal elements show strong exponential suppression under local observables. Regarding the strong ETH criteria, while the maximal diagonal elements exhibit systematic exponential decay with system size \(N\), the maximal off-diagonal elements show fluctuating behavior, albeit with moderate exponential suppression.
    \item \textbf{Integrable systems:} In contrast, integrable systems show a different pattern. While the average off-diagonal elements display significant exponential suppression for local observables, the suppression of average diagonal elements is noticeably weaker compared to non-integrable systems. In the strong ETH context, the maximal diagonal elements remain almost independent of \(N\), while the maximal off-diagonal elements show weaker exponential suppression, with reduced fluctuation amplitudes relative to non-integrable systems.
\end{itemize}
For spatially averaged observables, universal smoothing effects reduce the magnitudes of distinguishability while preserving key integrability-driven divergences:
\begin{itemize}
    \item \textbf{Non-integrable systems:} Under weak ETH, spatially averaged observables show a mildly accelerated off-diagonal decay, but slower diagonal suppression compared to local measurements. The average diagonal elements still exhibit significant exponential suppression, which is sufficient to confirm weak ETH. However, for strong ETH, both the maximum diagonal and off-diagonal elements show reduced exponential suppression rates.
    \item \textbf{Integrable systems:} The suppression trends for integrable systems are similar to those in non-integrable cases. However, average diagonal elements decay even more slowly compared to local measurements. For strong ETH, while the maximal diagonal elements show a weak suppression trend, their overall suppression remains insufficient. And the maximal off-diagonal elements become almost entirely unsuppressed.
\end{itemize}

% Our investigation concentrates on energy shells \(\mathcal{M}'_{E=0,\Delta}\) and \(\mathcal{M}_{E(\beta),\Delta}\) (\(E(\beta)\) determined from the canonical ensemble \(\rho^\text{c}(\beta=0.1)\)), revealing pronounced integrability-dependent signatures of strong ETH compliance according to \cref{ETHLM,ETHAO}: In non-integrable systems, maximum diagonal elements display clear \(N\)-dependent exponential suppression for local observables, while off-diagonal maxima exhibit fluctuating yet suppressed behavior—consistent with approximate strong ETH adherence. Integrable systems, however, violate strong ETH entirely due to the absence of \(N\)-dependent diagonal decay. For spatially averaged observables, smoothing effects reduce distinguishability magnitudes universally, yet integrability distinctions persist: non-integrable systems show slightly accelerated off-diagonal decay but retarded diagonal suppression compared to local measurements, whereas integrable systems retain largely \(N\)-independent off-diagonal maxima and sluggish diagonal suppression. 

The preceding analysis addressed fundamental eigenstate thermalization properties. We now turn to the parameter \(f^i\), which holds direct relevance to this work's central theme. Under the conventional ETH \cref{CMETH}, diagonal and off-diagonal matrix elements are generally assumed to share comparable scales, implying \(f^i \sim \mathcal{O}(1)\). However, as discussed in \cref{LOCase}, the system-size (\(N\)) dependence of \(f^i\) together with trade-off relation critically determines ETH validity, making its behavior pivotal. We investigate \(f^i\) in model \cref{HMT} by examining two energy shells: \(\mathcal{M}'\) and \(\mathcal{M}\). For \(\mathcal{M}\), we consider both the canonical ensemble \(\rho^\text{c}(\beta=0.1)\) and the microcanonical ensemble \(\rho^{\text{mc}} = \sum_{E_i \in \mathcal{M}} \Pi^i / d_{\mathcal{M}}\); for \(\mathcal{M}'\), we adopt the microcanonical ensemble \({\rho'}^\text{mc} = \sum_{E_i \in \mathcal{M}'} \Pi^i / d_{\mathcal{M}'}\).  

The ensemble-averaged \(f^i\) within an energy shell is computed as:  
\begin{align}
    \overline{f(N)}=\frac{1}{d_{\mathcal{M}_{E,\Delta}}}\sum_{E_i\in \mathcal{M}_{E,\Delta}} f^i(N).
\end{align} 
while the maximum value across states,  
\begin{equation}
    f_{\max}(N)= \max_{E_i\in\mathcal{M}_{E,\Delta}} f^i(N).
\end{equation}
quantifies whether \(f^i\) grows exponentially at a rate substantially slower than the Hilbert space dimension \(D_N\).  

As shown in \cref{FIGsfi}, integrable systems exhibit \(f^i\) growth rates nearly comparable to \(D_N\). According to \cref{DoDratio}, this will lead to ineffective or weak exponential suppression of diagonal elements — both in maxima and averages. In stark contrast, non-integrable systems display significantly slower \(f^i(N)\) scaling, resulting in strong exponential suppression of diagonal elements for both maximum and averaged values. This dichotomy underscores \(f^i\)'s role as a diagnostic marker for ETH compliance, bridging microscopic dynamics and macroscopic thermalization thresholds.

\section{Conclusion and discussion}\label{CD}
In this work, we establish a universal framework for probing eigenstate thermalization through a variance-based measure  $d^{(3)}$, anchored by a fundamental trade-off between diagonal elements $\{V_\text{dg}^{i}\}$ and the off-diagonal elements $\{V_\text{off}^{ij}\}$. The trade-off equation provides a model-agnostic constraint that unifies diagonal and off-diagonal ETH elements. Unlike trace-norm-based measure, this equation explicitly quantifies how exponentially many off-diagonal terms enforce suppression of average off-diagonal elements, irrespective of integrability. Crucially, $d^{(3)}$ retains analytical tractability while eliminating observable dependence—bridging subsystem ETH, weak ETH, and macroscopic observables under a single formalism. Our numerical analysis of the Ising chain reveals that integrability controls the growth rate of $f^i(N)$, a ratio determining whether diagonal elements evade exponential suppression. In non-integrable systems, chaotic dynamics drive $f_{\max} \ll D_N$, ensuring that all diagonal elements decay rapidly ($ V^\text{Loc}_{\text{dg},\max} \sim 2^{-0.51N}$ for PBC and $\sim 2^{-0.30N}$ for OBC), thus making the strong ETH tenable. In contrast, integrable systems exhibit $f_{\max} \sim D_N$, which preserves the local distinguishability of certain diagonal elements that hardly decays with the size of the system. This makes the strong ETH violated. Even if only the average distinguishability is considered, due to the fact that $\overline{f} \sim (D_N)^{0.78}$ for PBC and $\overline{f} \sim (D_N)^{0.88}$ for OBC, the decay rate of the diagonal element will be greatly slowed down  (Fig. \ref{FIGs3}b).

While $f^i(N)$ emerges as a pivotal diagnostic for ETH compliance (Eq. \ref{DoDratio}), our current analysis leaves open how microscopic interactions sculpt its scaling. Specifically, the interplay between integrability-breaking terms (e.g., longitudinal field $h$) and multi-body correlations may reconfigure the competition between diagonal and off-diagonal elements. Recent advances in foundational ETH theory \cite{HHTG24} demonstrate that the Gaussian eigenvalue statistics of interacting systems—combined with Dyson Brownian motion—suffice to derive ETH as an emergent property. This is a promising path to analytically link $f^i(N)$ to parameters like interaction range and strength. Future studies will adopt this methodology to dissect how integrability transitions (e.g., tuning $h$ from $0$ to $0.1$) dynamically reshape $f^i(N)$, potentially revealing universality classes beyond the Ising paradigm.

% The diagonal element's significance lies in its ratio to the average off-diagonal elements, which relates to the system's integrability. We also establish the trade-off relationship for weak ETH and average observables between their respective diagonal and off-diagonal elements. Our numerical results demonstrate that the measure $V(\rho_B,O^{B_1}_{ij})$ is indeed effective for analyzing ETH, and the strict exponential suppression of the average off-diagonal element aligns with the trade-off relationship's predictions.

%%%%%%%%%%%%这里主要讨论的是子系统ETH，对于考虑演化的ETH，可积模型通常将GGE作为稳态。

In the numerical calculations presented in this paper, we compare eigenstates within an energy shell using the same ensemble. However, the energies of these states may deviate from that of the ensemble, introducing additional distinguishability, as seen in \cref{FIGs2}(a). By using an ensemble with an energy that exactly matches the energy eigenstates, we can further reduce both the average and maximum distinguishability. Additionally, examining how the energy width of the microcanonical ensemble and interactions influence the ratio between diagonal and average off-diagonal elements is crucial for deepening our understanding of the theoretical mechanisms behind the ETH.

% The difficulty lies in how to obtain the corresponding microcanonical ensemble based on the eigenstate energy and quickly obtain its subsystem state. When $N$ is large, the number of states within the energy shell is large. Constructing a microcanonical ensemble and obtaining the state of its subsystem is a relatively time-consuming operation.

Additionally, this paper does not delve into cases where the size of subsystem $B_1$ scales proportionally with the full system,  $N_{B_1}/N_{B}=p$, as this has already been thoroughly discussed in \cite{DLL18}. However, using the measure form $d^{(3)}$ from this paper, there are still aspects worth exploring. For local observations, since  $\Tr(\rho_{B_1}^{-1})-1$ generally increases with $D_{B_1}^2$, if the ratio of diagonal to average off-diagonal elements $f^i(N)$ increases exponentially at a much slower rate than $D_N$, then we would have $ V_\text{dg}^{i}\propto D_{B_1}^2/D_N$. According to \cref{d2d3r}, we obtain $d^{(2)}\sim D_{B_1}/D_N^{1/2}$, which aligns with the results in \cite{DLL18}. Similarly to \cref{DGODGR}, we find that
\begin{align}
    V(\rho^{\text{c}}_B,O^{B_1}_{ii})=\sum_{\alpha,\beta}  {p'_\beta}^{-1}\abs{\bra{\beta}(\sigma_{B_1}^{i}-\rho_{B_1}^{\text{c}})\ket{\alpha}}^2 \notag\\
    \leq \sum_{\alpha,\beta}  {p'_\beta}^{-1}(\abs{\bra{\beta}(\sigma_{B_1}^{i}-\rho_{B_1}^{\text{mc}})\ket{\alpha}}+\abs{\bra{\beta}(\rho_{B_1}^{\text{mc}}-\rho_{B_1}^{\text{c}})\ket{\alpha}})^2. 
\end{align}
When $B_1$ is scaled to infinity, a semiclassical analysis similar to that in \cite{DLL18} can be performed to approximate $V(\rho^{\text{c}}_B,O^{B_1}_{ii})$. This provides an alternative perspective for considering the local differences between energy eigenstates and the canonical ensemble in the thermodynamic limit.

While our approach offers an alternative measures for traditional ETH, it inherits the latter's limitations in addressing energy degeneracy. In systems with exact or near-degenerate eigenstates, the conventional ETH framework may fail to account for potential variations in local observables across degenerate subspaces.  A critical direction for future work is to generalize our variance measure by partitioning the Hilbert space into symmetry-protected subspaces \cite{F25} or adopting the generalized ETH (GETH) framework \cite{WTP22,MBISY23}. Specifically, applying our trade-off relation to each symmetry-resolved subspace (e.g., conserved charge sectors) could reconcile the suppression of off-diagonal terms with degenerate diagonal elements. Furthermore, proving analogous variance bounds for GETH—where thermalization occurs within subspaces defined by additional conserved quantities—would extend our results to integrable and symmetry-rich systems. Such extensions may also clarify how degeneracy-induced effects constrain thermalization, bridging the gap between ETH and generalized Gibbs ensembles. These efforts will deepen the understanding of thermalization mechanisms in systems beyond the traditional chaotic paradigm.
\begin{acknowledgments}
We would like to thank Xiao-Kan Guo for his helpful comments.
This work is supported by the National Natural Science Foundation of
China under Grant No. 12305035. 
\end{acknowledgments}

\appendix
\section{Variance relation}\label{A1}
Here, we present the detailed derivation of the key quantum variance relation given in \cref{DGODGRM,avgotdoff} of the main text.  Starting with the spectral decomposition of the reduced density matrix $\rho_{B_1}$ within its Hilbert space framework, this operator admits the standard representation
\begin{equation}\label{specdc}
    \rho_{B_1}=\sum_\alpha p'_\alpha \Pi^\alpha_{B_1},
\end{equation}
where $\{\Pi^\alpha_{B_1}\}$ constitutes a complete set of orthogonal projection operators. This spectral structure allows us to reformulate \cref{VCVF} as 
\begin{align}\label{DGODG}
    V_\text{dg}^{i}+ \sum_{j\neq i} V_\text{off}^{ij}= \sum_{j,\alpha,\beta}  {p'_\alpha}^{-1}\abs{\bra{\beta}\sigma^{ij}_{B_1}\ket{\alpha}}^2-1\notag\\
    = \sum_{j,\alpha,\beta}  {p'_\alpha}^{-1}\abs{\Tr(\sigma^{ij}_{B_1}\Pi^{\alpha \beta}_{B_1})}^2-1,
\end{align}
where the elementary operators $\Pi^{\alpha \beta}_{B_1}=\ket{\alpha }_{B_1}\! \bra{\beta}$ span the operator space of the subsystem, forming a complete basis for operator expansion. Recognizing that $\abs{\Tr (\sigma^{ij}_{B_1}  A^{B_1})}^2=\abs{\bra{E_j}  A^{B_1}\ket{E_i}}^2$, we can further rewrite
\begin{align}\label{DGODGR}
    V_\text{dg}^{i}+ \sum_{j\neq i} V_\text{off}^{ij}= \sum_{j,\alpha,\beta}  {p'_\alpha}^{-1}\abs{\bra{E_j}\Pi^{\alpha \beta}_{B_1}\ket{E_i}}^2-1\notag\\
   = \sum_{\alpha,\beta}  {p'_\alpha}^{-1}\bra{E_i}\Pi^{ \beta\alpha}_{B_1}\Pi^{\alpha \beta}_{B_1}\ket{E_i}-1 \notag\\
   =\sum_{\alpha}  {p'_\alpha}^{-1}\bra{E_i}I_{B_1}\ket{E_i}-1=\Tr(\rho_{B_1}^{-1})-1,
\end{align}
where in the second equality we use the orthogonal projection completeness condition $\sum_j  \ket{E_j} \bra{E_j}=I$.
The \cref{DGODGR} directly corresponds to the relation in \cref{DGODGRM}. Another derivation method that is easy to generalize to average observable needs to use the following relationship from \cref{specdc}
\begin{equation}
    \mathcal{J}_{\rho_{B_k}}^{-1/2} (\Pi^{\alpha \beta}_{B_k})=(p'_\alpha p'_\beta)^{-1/2}\Pi^{\alpha \beta}_{B_k},
\end{equation}
along with the definition
\begin{equation}
    O^{B_1}_{\alpha \beta}=\mathcal{J}_{\rho_{B_1}}^{-1/2} (\Pi^{\alpha \beta}_{B_1}).
\end{equation}
Using these two equations and $O^{B_1}_{\beta \alpha}=(O^{B_1}_{\alpha \beta})^\dag$, we can prove that
 \begin{align}\label{DGODGR2}
    V_\text{dg}^{i}+ \sum_{j\neq i} V_\text{off}^{ij}= \sum_{j,\alpha,\beta}  {p'_\alpha}^{-1}\abs{\bra{E_j}\Pi^{\alpha \beta}_{B_1}\ket{E_i}}^2-1\notag\\
   =\sum_{\beta, \alpha} p'_\beta \Tr [O^{B_1}_{\alpha \beta}\Pi^i_B (O^{B_1}_{\alpha \beta})^\dag]-1 \notag\\
   =\sum_{\beta, \alpha} p'_\alpha \Tr [(O^{B_1}_{\alpha \beta})^\dag \Pi^i_B O^{B_1}_{\alpha \beta}]-1 \notag\\
   =\sum_{\beta, \alpha}  {p'_\beta}^{-1} \Tr (\Pi^{\beta \alpha}_{B_1}\Pi^i_B\Pi^{\alpha \beta}_{B_1})-1=\Tr(\rho_{B_1}^{-1})-1.
\end{align}

The proof of \cref{avgotdoff} follows similar approach as \cref{DGODGR2}. Different from the $ V(\rho_B,O^{B_1}_{ij})$ considered in \cref{DGODGR2}, what we need to consider here is $ V(\rho_B,\overline{O}^{B_1}_{ij})$ with $\overline{O}^{B_1}_{ij}=\frac{1}{\mathcal{C}}\sum_{k}\mathcal{J}_{\rho_{B_1}}^{-1/2} (\sigma^{ij,k}_{B_1})$. The key difference is the substitution of $O$ with $\overline{O}$. Following \cref{DGODG} and using the relation
\begin{equation}\label{SDTIOFF}
    \abs{\Tr [(\frac{1}{\mathcal{C}}\sum_k\sigma^{ij,k}_{B_1}) \Pi^{\alpha \beta}_{B_1}]}^2=\abs{\Tr [\Pi^{ij} (\frac{1}{\mathcal{C}}\sum_k \Pi^{\alpha \beta,1}_{B_k})]}^2  
 \end{equation}
with $\Pi^{\alpha \beta,1}_{B_k}$ represents the translational copy of $\Pi^{\alpha \beta}_{B_1}$, we can reach a formula similar in \cref{DGODGR2}
\begin{align}\label{SIMDGODGR2}
    V(\rho_B,\overline{O}^{B_1}_{ii})+\sum_{j\neq i} V(\rho_B,\overline{O}^{B_1}_{ij})\notag\\
    =\sum_{\beta, \alpha} p'_\alpha \Tr [(\overline{O}^{B}_{\alpha \beta})^\dag\Pi^i_B \overline{O}^{B}_{\alpha \beta}]-1, 
\end{align}
where $\overline{O}^{B}_{\alpha \beta}=\frac{1}{\mathcal{C}}\sum_k O^{B_k}_{\alpha \beta,1}$ and $O^{B_k}_{\alpha \beta,1}=\mathcal{J}_{\rho^1_{B_k}}^{-1/2} (\Pi^{\alpha \beta,1}_{B_k})$ represents the translational copies of $O^{B_1}_{\alpha \beta}$. 
Go on with \cref{SIMDGODGR2}, we have
\begin{align}\label{avgoDGODGR}
    V(\rho_B,\overline{O}^{B_1}_{ii})+\sum_{j\neq i} V(\rho_B,\overline{O}^{B_1}_{ij})\notag\\
    =\frac{1}{\mathcal{C}^2}\sum_{\beta, \alpha,k} p'_\alpha \Tr [(O^{B_k}_{\alpha \beta,1})^\dag\Pi^i_B O^{B_k}_{\alpha \beta,1}]-\frac{1}{\mathcal{C}}\notag\\
    +\frac{1}{\mathcal{C}^2}\sum_{\beta, \alpha,l,k\neq l} p'_\alpha \Tr [(O^{B_k}_{\alpha \beta,1})^\dag\Pi^i_B O^{B_l}_{\alpha \beta,1}]-\frac{\mathcal{C}-1}{\mathcal{C}}.
\end{align}
In this expression, the local terms are given by
\begin{align}
    \sum_{\beta, \alpha} p'_\alpha \Tr [(O^{B_k}_{\alpha \beta,1})^\dag\Pi^i_B O^{B_k}_{\alpha \beta,1}]\notag\\
    =\sum_{\beta}  {p'_\beta}^{-1} \Tr (\Pi^{\beta \alpha}_{B_k}\Pi^i_B\Pi^{\alpha \beta}_{B_k})=\Tr(\rho_{B_1}^{-1}).
\end{align}
Combining this with
\begin{equation}
    \Tr [(O^{B_k}_{\alpha \beta,1})^\dag \sigma^{ii}_{B_k}]\times\Tr(\rho^{1}_{B_l} O^{B_l}_{\alpha \beta,1})={p'_\alpha}^{-1}\Tr [\Pi_{\alpha}\sigma^{ii}_{B_k}]\delta_{\alpha\beta},
\end{equation}
we arrive at \cref{avgotdoff}.

\begin{figure*}
    \centering
    \subfigure[]{\includegraphics[width=0.48\textwidth]{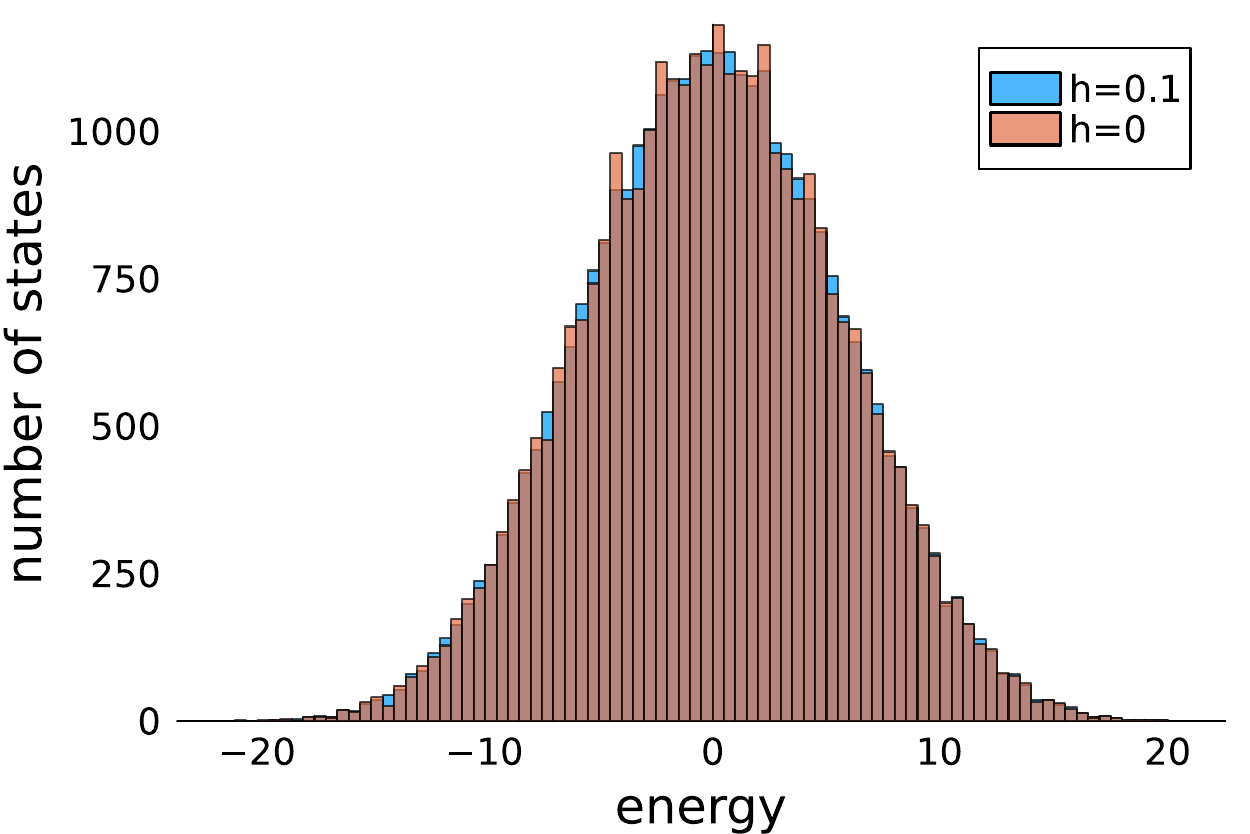}} 
    \subfigure[]{\includegraphics[width=0.48\textwidth]{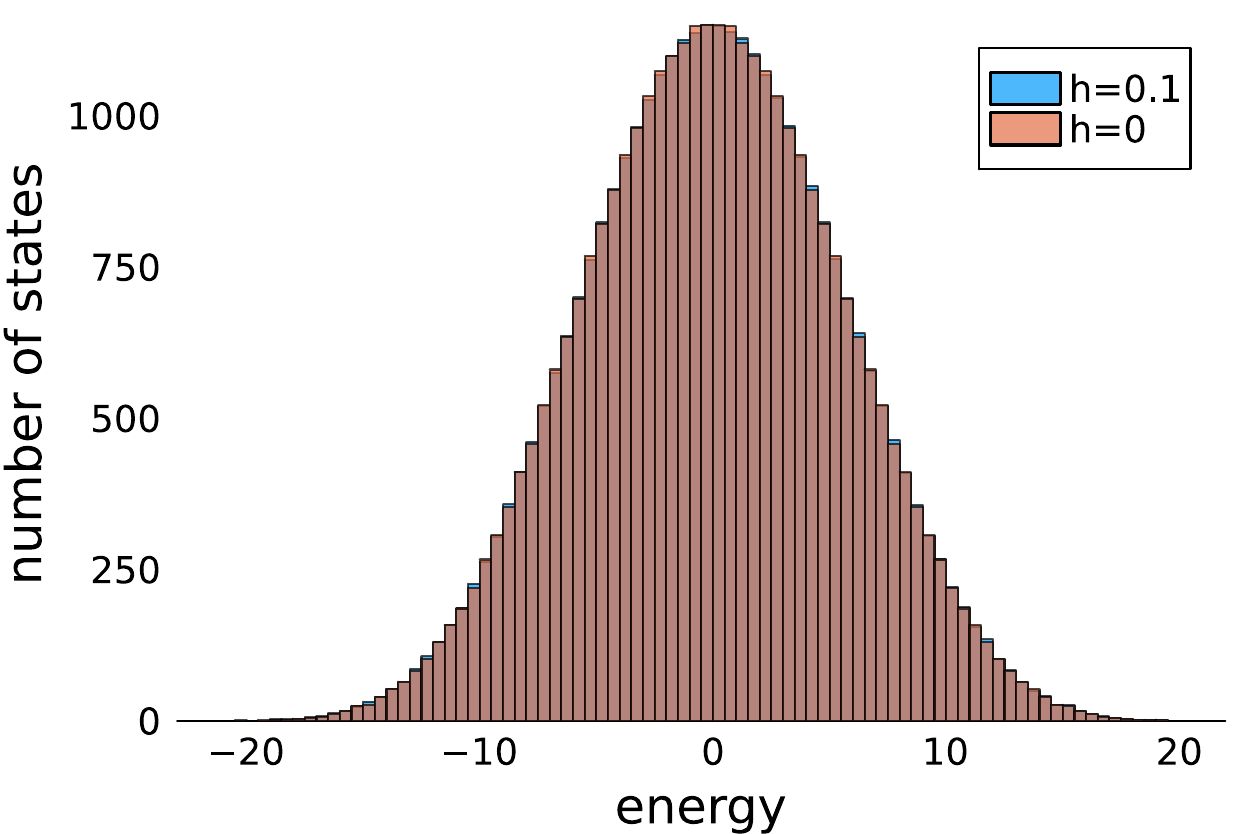}} 
    \subfigure[]{\includegraphics[width=0.48\textwidth]{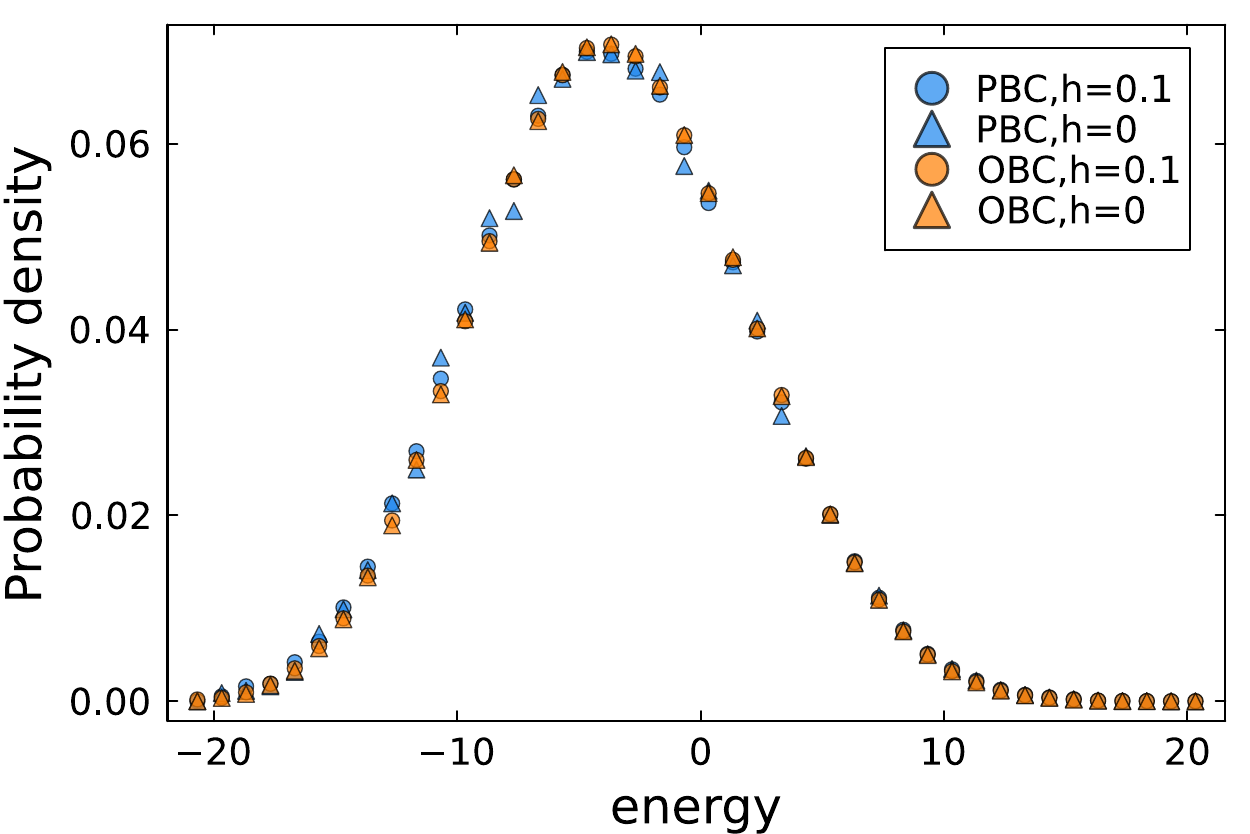}} 
    \subfigure[]{\includegraphics[width=0.5\textwidth]{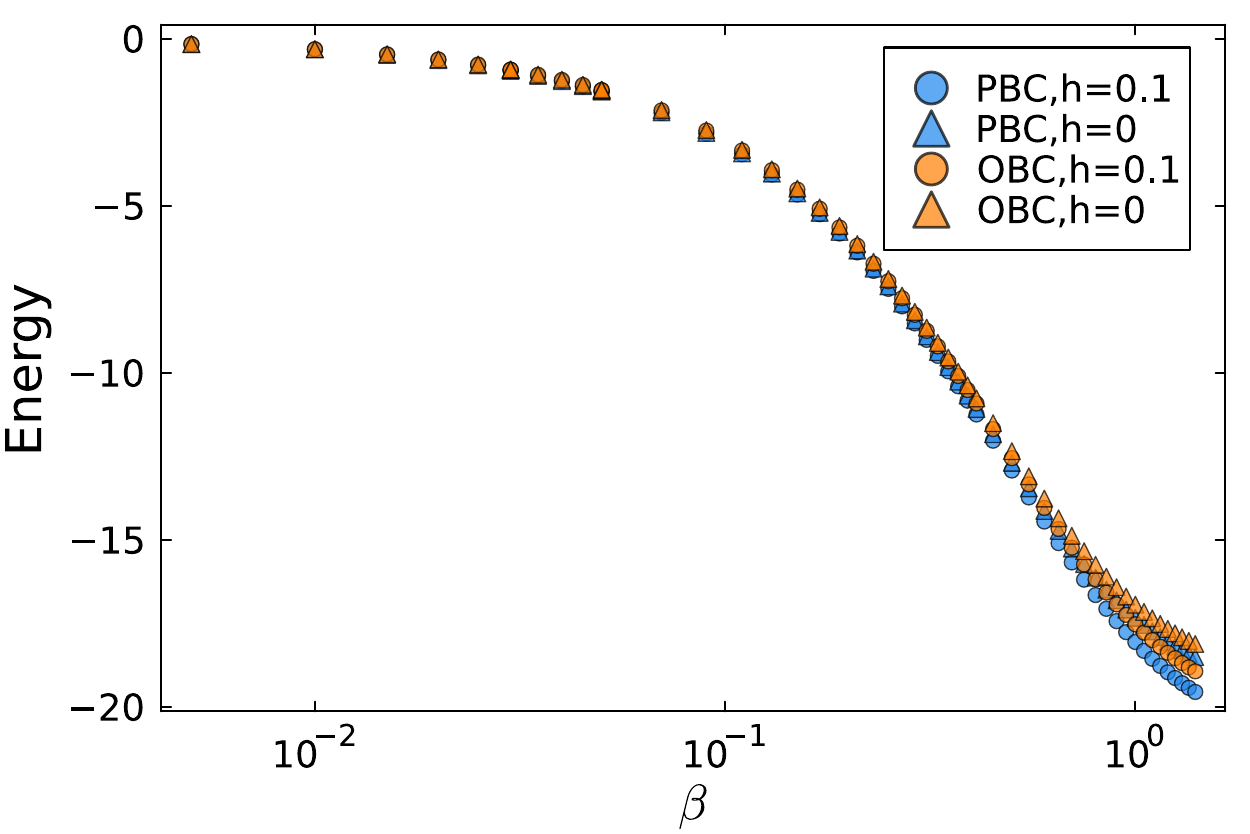}} 
    \subfigure[]{\includegraphics[width=0.48\textwidth]{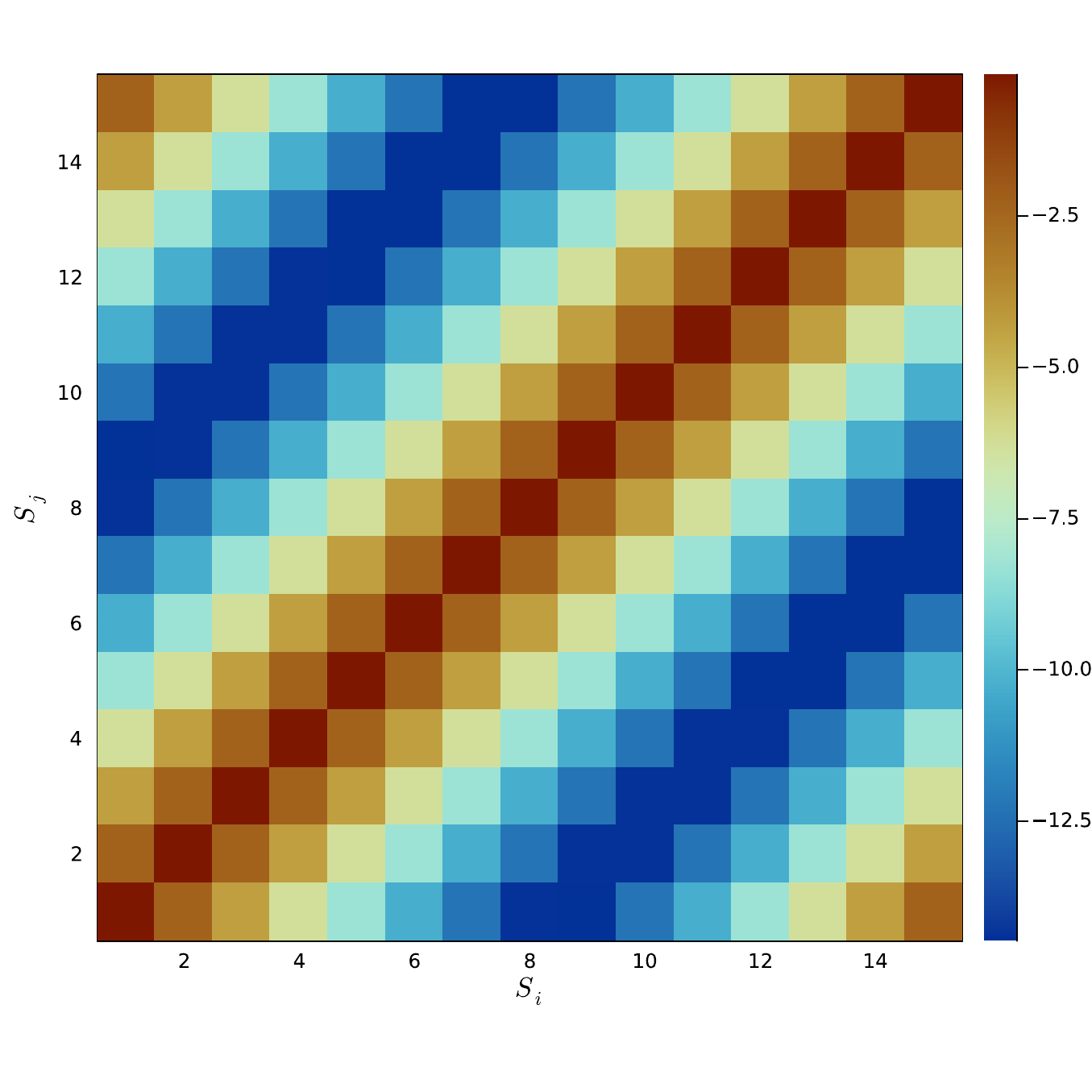}} 
    \subfigure[]{\includegraphics[width=0.48\textwidth]{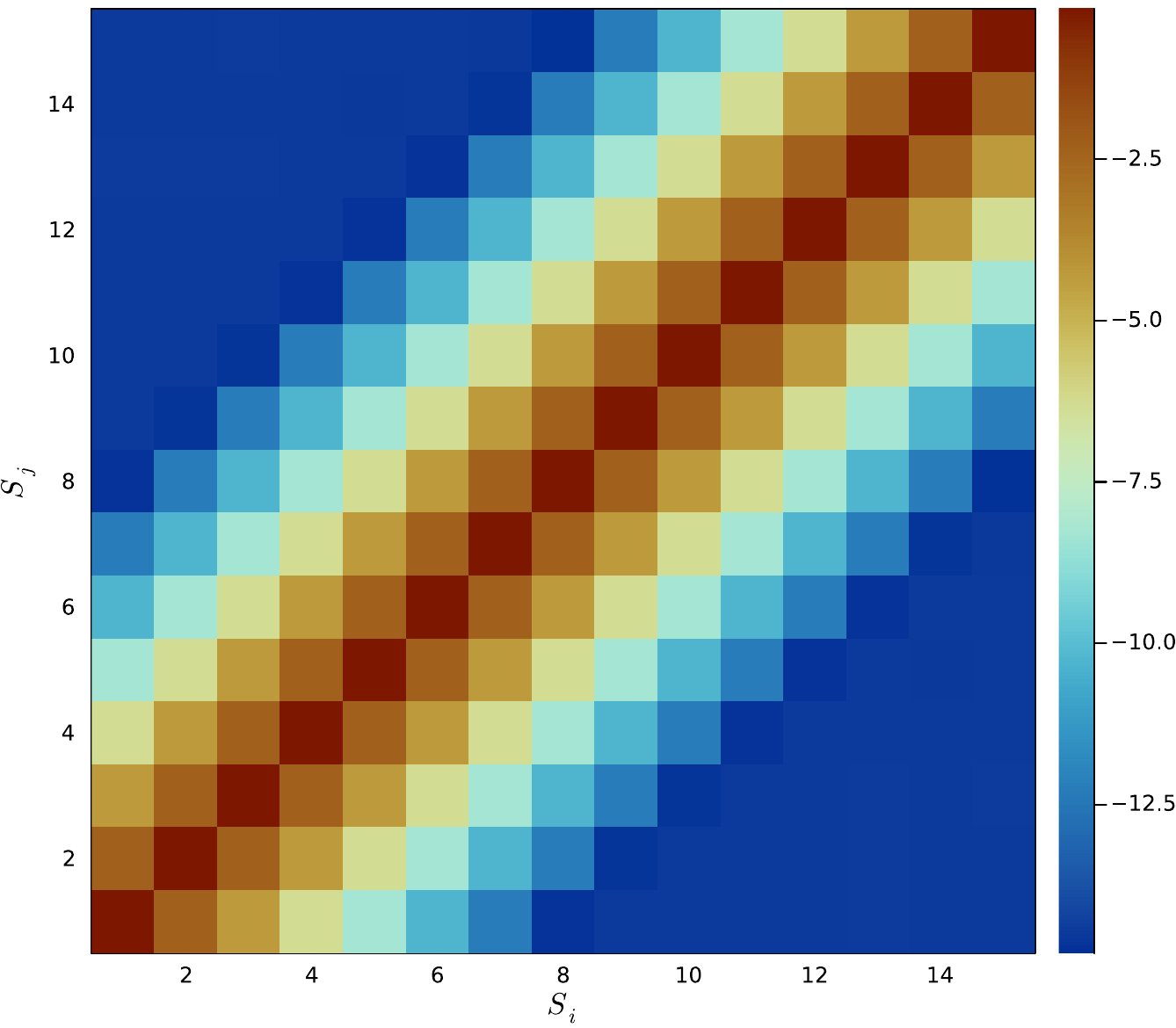}} 
    \caption{Parameters: $g =
    1.05$,  $N = 15$. (a-b) The number of states of in a spin chain under PBC (OBC).  (c)  Energy probability distribution of the canonical ensemble with $\beta =0.1$. (d) Energy as a function of the thermodynamic beta. (e-f) The value of $\log_{10}I(S_i,S_j)$ between two sites $S_i$ and $S_j$ at $\beta =0.1$ in a spin chain under PBC (OBC).}
    \label{FIGs}
\end{figure*}
% \section{Another relation between diagonal and non-diagonal elements}
% 类似\cref{DGODGR}，我们这里推导另一个对角元与非对角元的关系，它提供了一个分析$f^i(N)$的角度。对于任意非负数集合$\{\gamma_j\}$,容易证明
% \begin{align}
%     \sum_j  V(\rho_B,O^{B_1}_{ij})\gamma_j=\sum_j\Tr(\sigma^{ij}_{B_1}\gamma_j^{\frac{1}{2}}\rho^{-1}_{B_1}\gamma_j^{\frac{1}{2}}\sigma^{ij\dag}_{B_1})-\gamma_i\notag\\
%     =\Tr(\gamma\rho_{B_1}^{-1})-\gamma_i
% \end{align}
% 上述证明是错的，根据A6最后一步,证明过不去。
% 其中$\gamma=\sum_j \gamma_j \Pi^j_B$.考虑$\gamma=\rho_B$的特殊情况，则有
% \begin{equation}
%     \sum_j  V(\rho_B,O^{B_1}_{ij})p_j=D_{B_1}-p_i
% \end{equation}
% 将它与\cref{DGODGRM}联立可得
% \begin{equation}
%     \sum_{j;j\neq i} V_\text{off}^{ij}(p_i-p_j)=p_i\Tr(\rho_{B_1}^{-1})-D_{B_1}
% \end{equation}
% 当$\rho_B$是微正则系综时，对系综内的态$p_i-p_j=0$,系综外的态则有$p_i-p_j=1/d_\mathcal{M}$.

\section{Numerical Verification of ETH}\label{NVETH}

\subsection{Basic properties of the model}\label{BPM}
Here, we briefly illustrate the basic properties of the model. As shown in \cref{FIGs}(a), the density of states for both integrable and non-integrable systems is quite similar, with the non-integrable system exhibiting a smoother distribution that can be reasonably approximated by a binomial distribution \cite{DLL18}. The energy-temperature curve for the system is provided in \cref{FIGs}(c). Additionally, as shown in \cref{FIGs}(d), the mutual information $I(B_1,B_2)$ decays exponentially with the distance between sites. The plot depicts the system with $h=0.1$; the mutual information for the $h=0$ system is nearly identical and therefore not shown. From the perspective of these basic properties, integrability has no obvious impact on them.

\subsection{The impact of energy difference and temperature on distinguishability}\label{IEDTD}
\begin{figure*}
    \centering
    \subfigure[]{\includegraphics[width=0.48\textwidth]{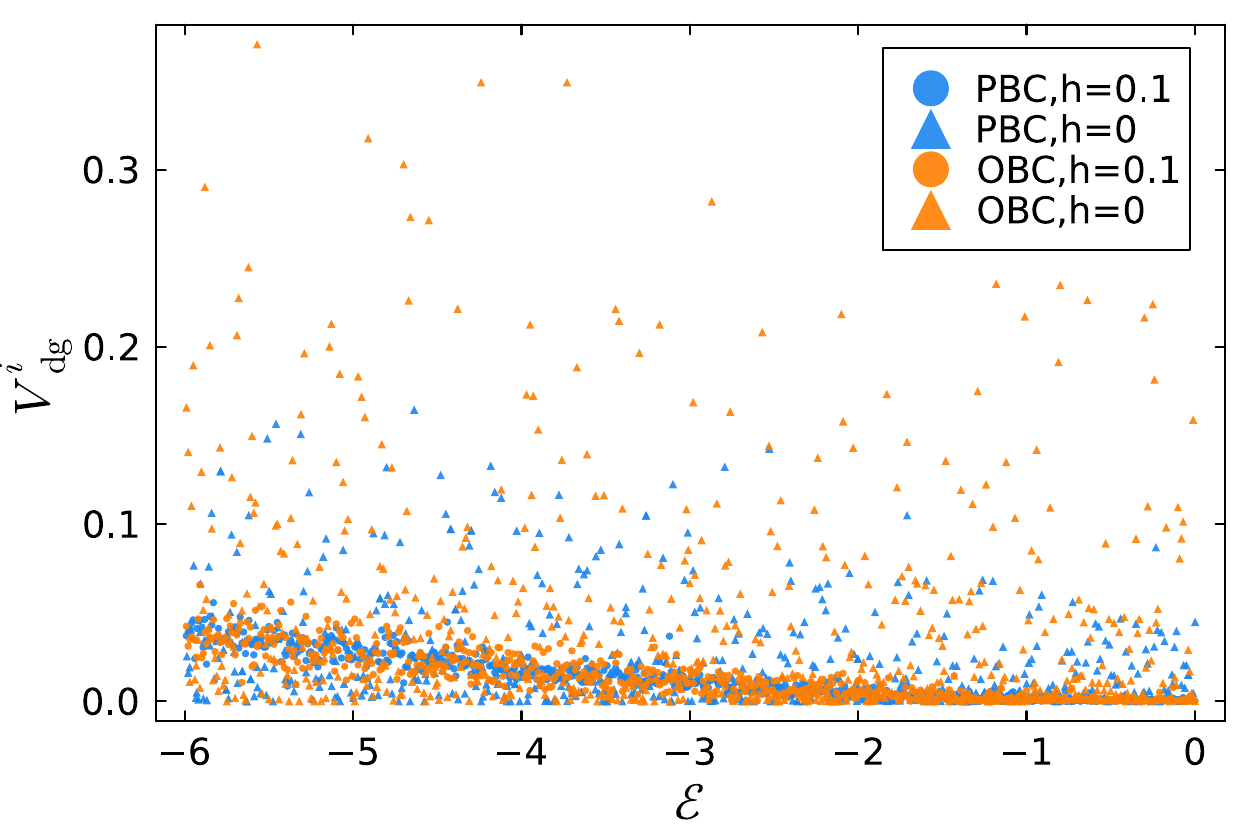}} 
    \subfigure[]{\includegraphics[width=0.48\textwidth]{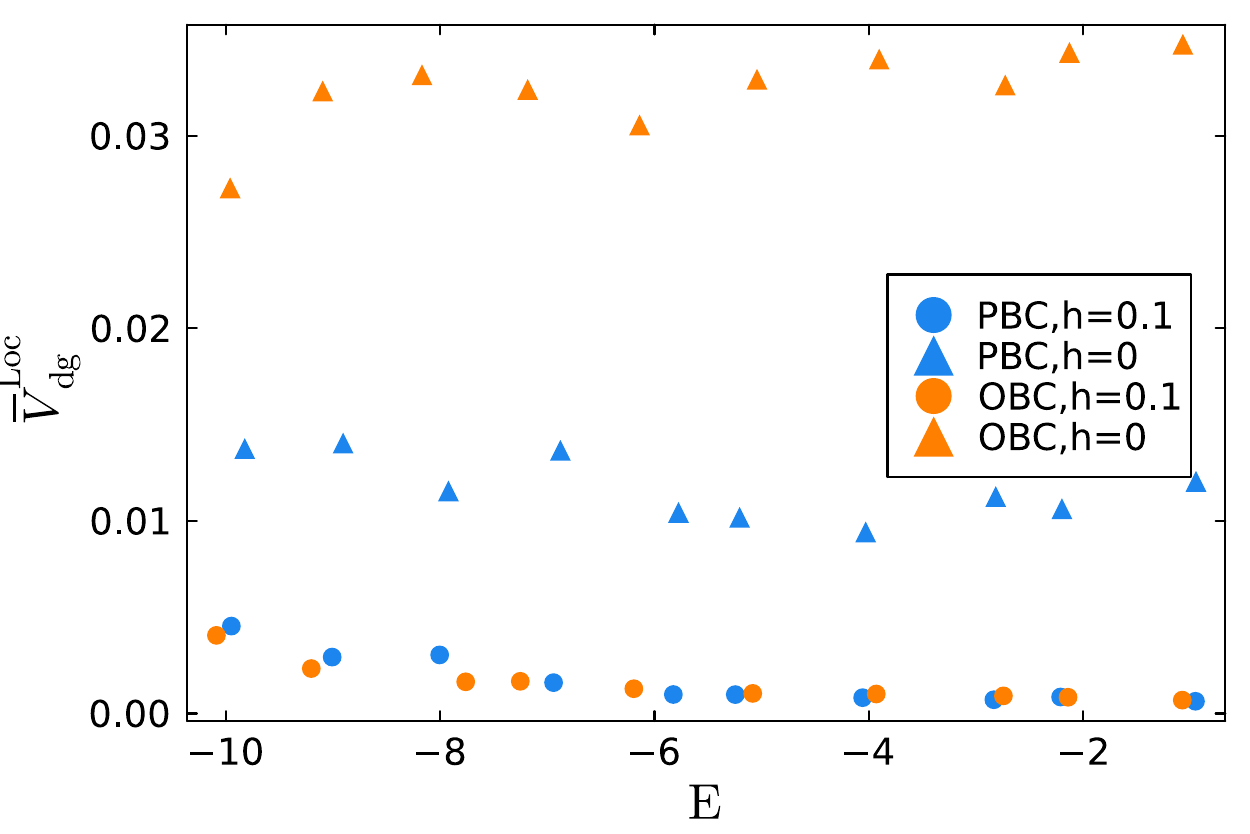}} 
    \caption{Parameters: $g =1.05$,  $N = 15$, $N_{B_1}=1$. (a) Local diagonal element measure $V_\text{dg}^{i}$ as a function of energy deviation $\mathcal{E} := E_i - E(\rho_B)$, where $\rho_B$ is fixed as a maximally mixed state $\rho_B = \sum_i \Pi^i_B/D_N$ with $E(\rho_B) = 0$.  For each situation, only a subset of $600$ states with approximately uniform energy spacing is plotted, rather than all states across the energy range.  (b) The average distinguishability $\overline{V}^\text{Loc}_\text{dg}$ within the energy shell $\mathcal{M}_{E,\Delta=0.2}$. For each energy shell centered at $E$,  the $\rho_B$ in $V_\text{dg}^{i}=V(\rho_B,O^{B_1}_{ii})$  is adaptively parametrized as $\rho^\text{c}(\beta)$ to satisfy $\Tr[H \rho^\text{c}(\beta)] = E$.}
    \label{FIGs2}
\end{figure*}

Although equation \eqref{DGODGRM} holds for all energy eigenstates, the measure $V_\text{dg}^{i}$ for the diagonal elements varies across different energy eigenstates $\ket{E_i}$. This difference is evident in \cref{FIG6}. In more detail, as shown in \cref{FIGs2}(a), there is some randomness in the measure $V_\text{dg}^{i}$ with respect to the eigenstate, but generally, the greater the energy difference between $E_i$ and $E(\rho_B)$, the higher the average distinguishability. In addition, integrability also has a great influence. For the integrable systems, the diagonal elements $V_\text{dg}^{i}$ tend to be more variable, and the average distinguishability is typically higher.

Given these factors, it is not always meaningful to consider the maximum value of 
$V_\text{dg}^{i}$ over the entire energy eigenstate space for a fixed $\rho_B$. Instead, it is more insightful to examine the maximum value within a specific energy shell, such as
\begin{equation}
    \mathcal{M}_{E,\Delta}=(E-\Delta /2,E+\Delta /2).
\end{equation}
Moreover, when analyzing energy eigenstates within $\mathcal{M}_{E,\Delta}$, it is preferable to use an ensemble $\rho_B$ with the same energy as $E$. In \cref{FIGs2}(b), we selected the canonical ensemble with corresponding energies for different energy shells and calculated the average distinguishability as  
\begin{equation}
    \overline{V}^\text{Loc}_\text{dg}= \sum_{E_i \in \mathcal{M}_{E,\Delta} }V_\text{dg}^{i}/d_{\mathcal{M}_{E,\Delta}},
\end{equation}
where $d_{\mathcal{M}}$ denotes the number of energy levels within the energy shell. Note that this average distinguishability is not the same as the ``diagonal'' probabilistic typicality in \cref{CLQV}. The $\rho_B$ used in $V_\text{dg}^{i}$ may not be the microcanonical ensemble consisting of states within $\mathcal{M}_{E,\Delta}$. According to \cref{FIGs2}(b), although the average distinguishability varies with the energy center $E$,  the variation is significantly less compared to the changes caused by energy differences, as shown in \cref{FIGs2}(a). Additionally, the average distinguishability in integrable systems remains higher than in non-integrable systems.

The results in \cref{FIGs2} are consistent with the traditional ETH: As the energy discrepancy between the ensemble and a particular energy eigenstate grows, their distinguishability increases correspondingly. Furthermore, the distinguishability can only be minimized when the ensemble is specifically tailored to match the energy of a given eigenstate.

\subsection{The ETH under the local measurement}\label{ETHLM}

In this section, we focus on analyzing the strong and weak ETH under local measurements. We examine two energy shells $ \mathcal{M}':=\mathcal{M}'_{E=0,\Delta}$ and  $\mathcal{M}:=\mathcal{M}_{E(\beta),\Delta}$, where $\Delta =0.2N$ and $E(\beta)=E(\rho^\text{c}(\beta=0.1))$. Both energy shells include highly excited states. For the energy shell $\mathcal{M}$, we considered the canonical ensemble $\rho^\text{c}(\beta=0.1)$ and the mirocanonical ensemble $\rho^{\text{mc}}= \sum_{E_i \in \mathcal{M} }\Pi^{i}/d_{\mathcal{M}}$. For the energy shell $\mathcal{M}'$, we used the mirocanonical ensemble ${\rho'}^\text{mc}= \sum_{E_i \in \mathcal{M}' }\Pi^{i}/d_{\mathcal{M}'}$. 

We compute the average distinguishability of the eigenstates within the energy shell as follows
\begin{align}\label{GavgLV}
    \overline{V}^\text{Loc}_\text{(off-)dg}=\frac{1}{d_{\mathcal{M}_{E,\Delta}}}\sum_{E_i\in \mathcal{M}_{E,\Delta}} V_\text{(off-)dg}^{i}.
\end{align}
For two microcanonical ensembles, this average corresponds to the  “(off-)diagonal” probabilistic typicality given in \cref{CLQVOFF} and \cref{CLQV}. To determine whether the diagonal elements $V_\text{dg}^{i}$ are exponentially small for all states in the energy shell, we calculate
\begin{equation}\label{LDGM}
    V^\text{Loc}_{\text{dg},\max}= \max_{E_i\in{\mathcal{M}_{E,\Delta}}} V_\text{dg}^{i}.
 \end{equation}
We also compute the maximum off-diagonal elements relative to the diagonal elements
\begin{equation}\label{LODGM}
    V^\text{Loc}_{\text{off},\max}= \max_{E_i\in{\mathcal{M}_{E,\Delta}};j} V_\text{off}^{ij}.
 \end{equation}
Here, as in \cref{DGODGRM}, the index $j$ spans the full Hilbert space.

\begin{figure}
    \centering
    \includegraphics[width=0.48\textwidth]{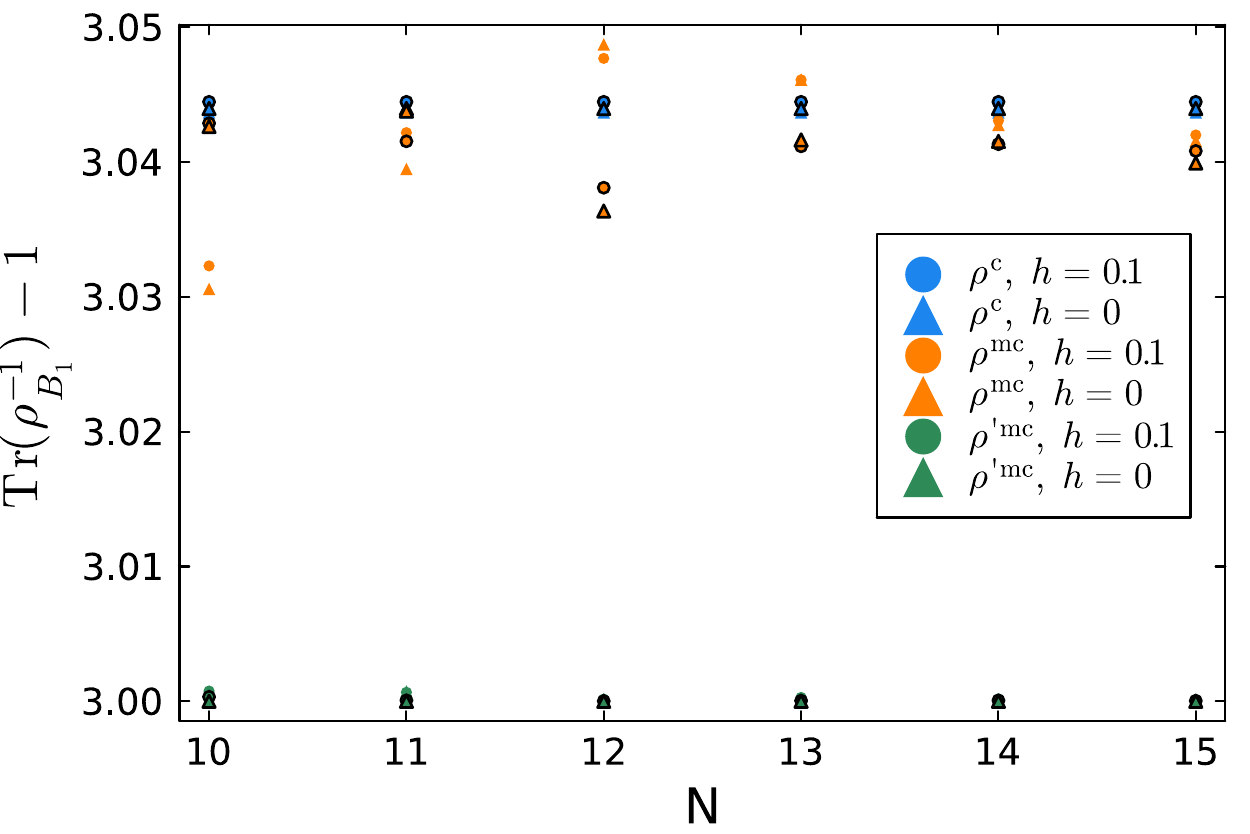}
    \caption{The local quantity $\Tr(\rho_{B_1}^{-1})-1$ under different system sizes. Parameters: $g =1.05$,  $N_{B_1}=1$. In the figure, zero markerstrokewidth corresponds to PBC, whereas non-zero markerstrokewidth indicates OBC.}\label{FIGloc}
\end{figure}

According to \cref{DGODGRM,LVVCAN}, the local state of ensemble $\rho_{B_1}$ is a crucial quantity. As $N$ increases from $10$ to $15$, according to \cref{FIGloc}, $\Tr(\rho_{B_1}^{-1})-1$ shows almost no variation across several ensembles, regardless of whether the system is integrable. 

\begin{figure*}
    \centering
    \subfigure[]{\includegraphics[width=0.48\textwidth]{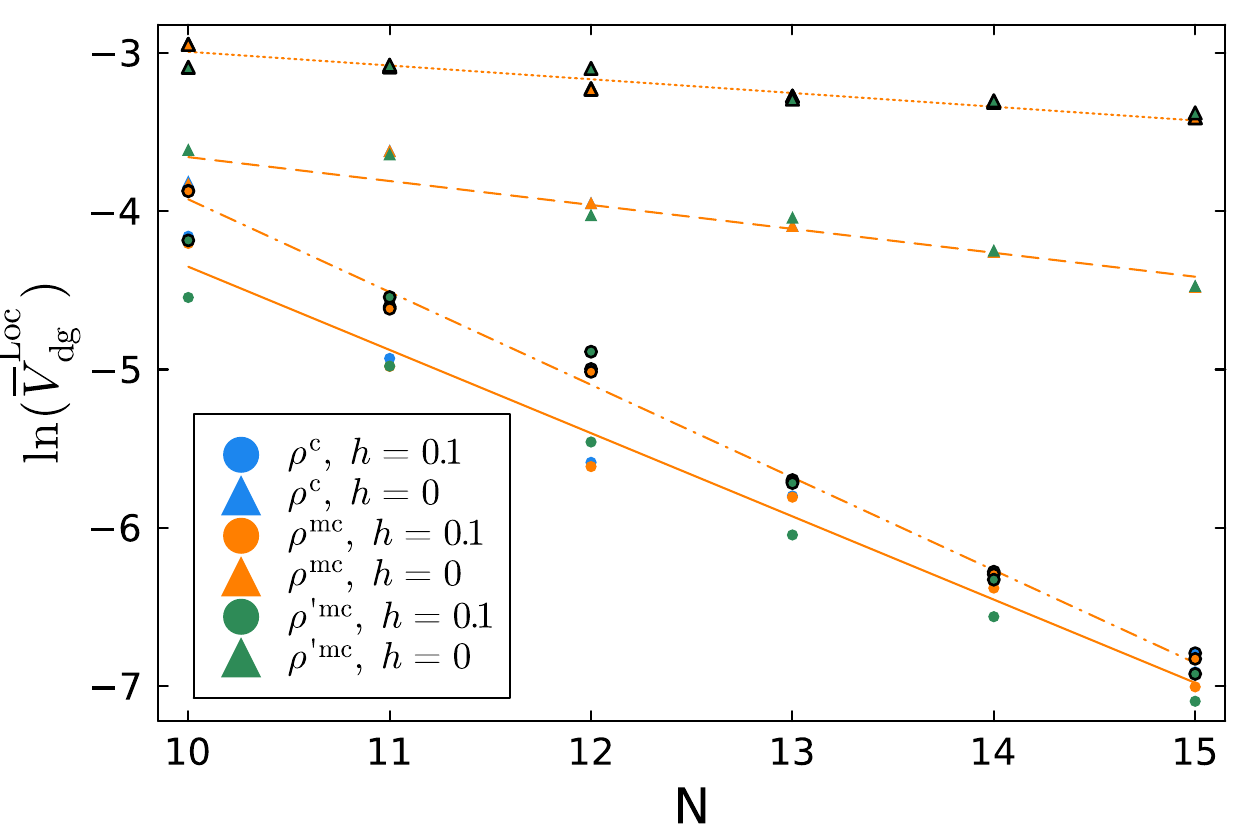}} 
    \subfigure[]{\includegraphics[width=0.48\textwidth]{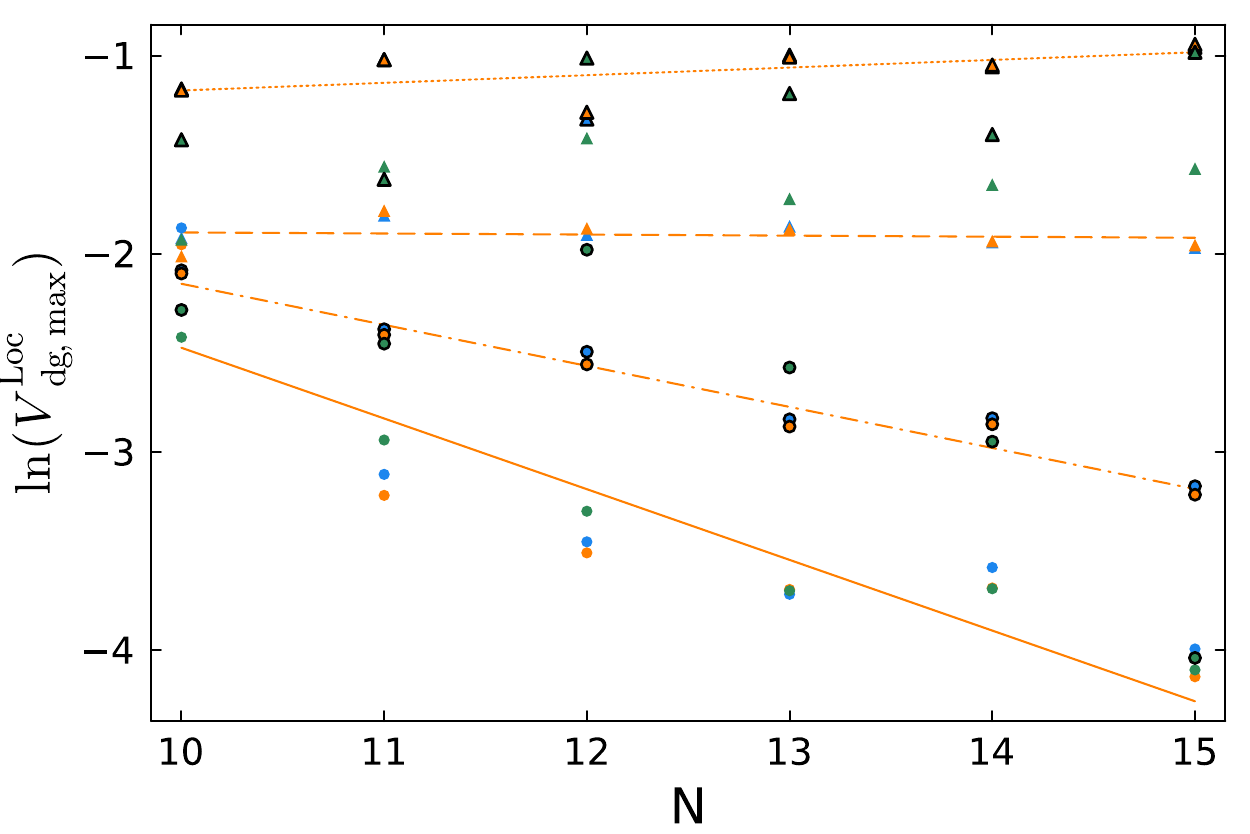}} 
    \subfigure[]{\includegraphics[width=0.48\textwidth]{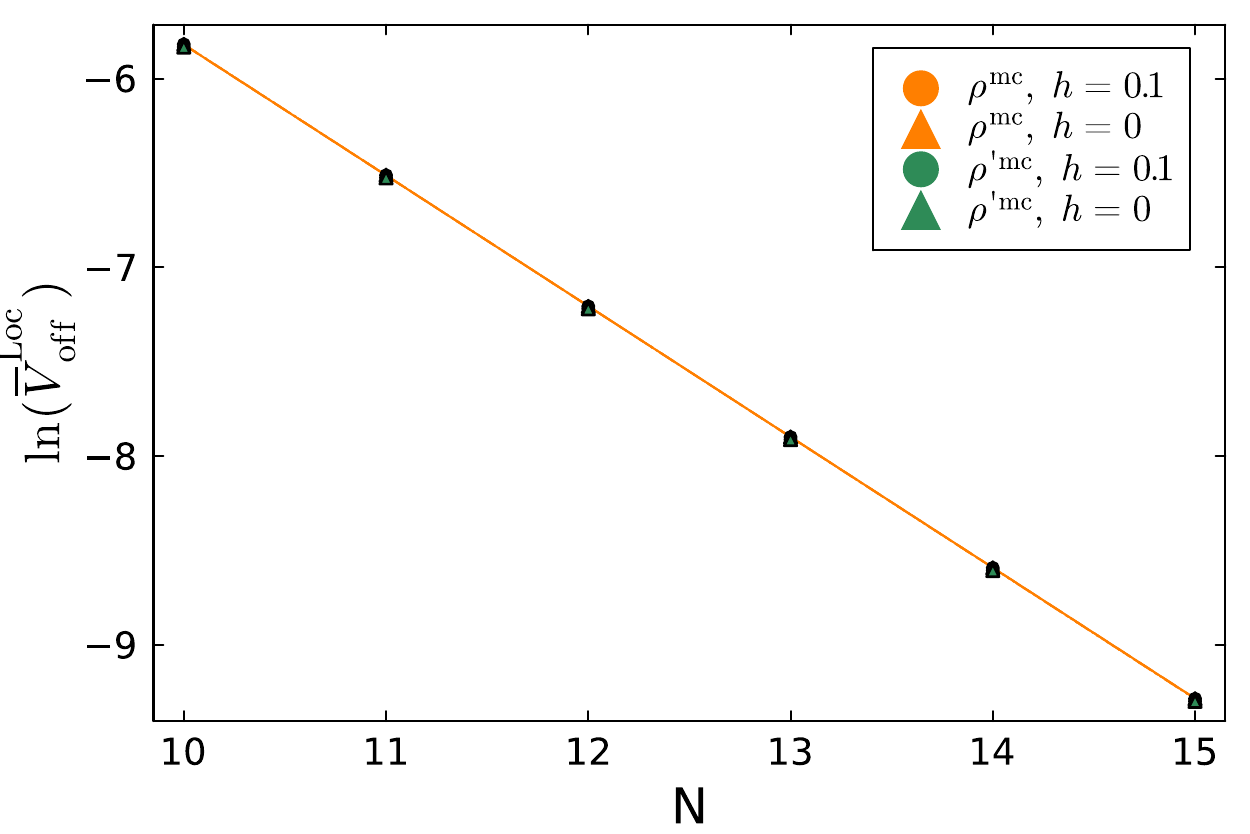}} 
    \subfigure[]{\includegraphics[width=0.48\textwidth]{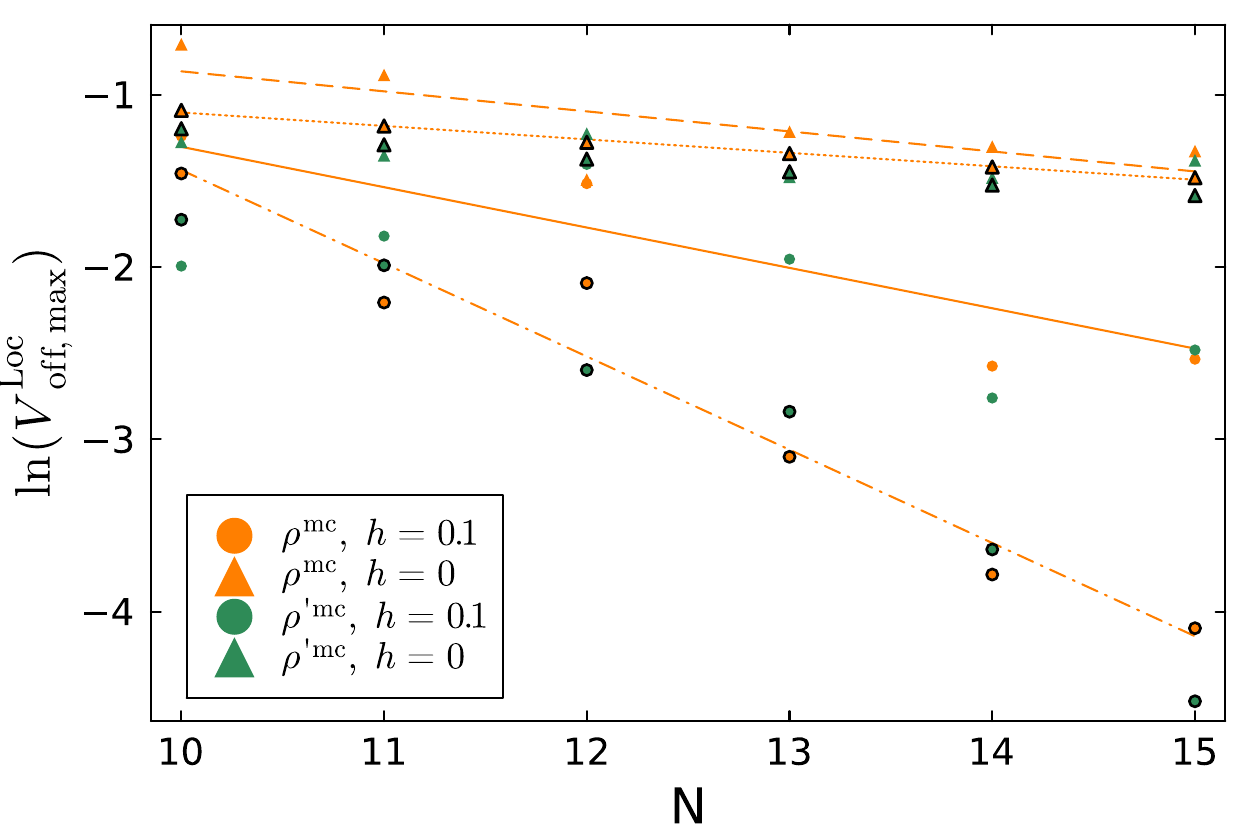}} 
    \caption{Parameters: $g = 1.05$,  $N_{B_1}=1$. In the figure, zero markerstrokewidth corresponds to PBC, whereas non-zero markerstrokewidth indicates OBC. Linear fits are shown exclusively for the energy shell $\mathcal{M}_{E(\beta),\Delta}$ with $\rho_B=\rho^{\text{mc}}$. The exponential decay constant is estimated using this linear fit.  (a) The average diagonal elements $\overline{V}^\text{Loc}_\text{dg}$ versus $N$. Exponential decay constants: Under PBC,  $0.76\times\ln 2$ (standard error $0.05\times\ln 2$) for the non-integrable system and $0.22\times\ln 2$ (standard error $0.05\times\ln 2$) for the integrable system.  Under OBC,  $0.84\times\ln 2$ (standard error $0.03\times\ln 2$) for the non-integrable system and $0.12\times\ln 2$ (standard error $0.02\times\ln 2$) for the integrable system.  (b) The maximum diagonal element $ V^\text{Loc}_{\text{dg},\max}$ versus $N$. Exponential decay constants: Under PBC,  $0.51\times\ln 2$ (standard error $0.13\times\ln 2$) for the non-integrable system and $0.00\times\ln 2$ (standard error $0.03\times\ln 2$) for the integrable system.  Under OBC,  $0.30\times\ln 2$ (standard error $0.03\times\ln 2$) for the non-integrable system and $-0.06\times\ln 2$ (standard error $0.04\times\ln 2$) for the integrable system.  (c) The average off-diagonal elements $\overline{V}^\text{Loc}_\text{off}$ versus $N$.  Both non-integrable and integrable systems exhibit similar exponential decay constants of $1.00\times\ln 2$ with standard errors $<0.001\times\ln 2$. (d) The maximum off-diagonal element $ V^\text{Loc}_{\text{off},\max}$ versus $N$. Exponential decay constants:  Under PBC,  $0.34\times\ln 2$ (standard error $0.14\times\ln 2$) for the non-integrable system and $0.17\times\ln 2$ (standard error $0.08\times\ln 2$) for the integrable system.  Under OBC,  $0.78\times\ln 2$ (standard error $0.09\times\ln 2$) for the non-integrable system and $0.11\times\ln 2$ (standard error $0.04\times\ln 2$) for the integrable system.
    }
    \label{FIGs3}
\end{figure*}

To avoid cluttering the figures, we perform linear fitting only for the microcanonical ensemble $\rho^{\text{mc}}$ with  energy shell $\mathcal{M}_{E(\beta),\Delta}$; all subsequent discussions of exponential decay constants are based on this case. As shown in \cref{FIGs3}(a-b),  there is a significant difference between integrable and non-integrable systems in the behavior of the diagonal elements. In the non-integrable system, the average distinguishability of the diagonal elements decreases exponentially with $N$.  The exponential decay constant is approximately $0.76\times\ln 2$ for PBC and $0.84\times\ln 2$ for OBC, which is closely aligns with the $\ln 2=d \ln(1/D_N)/d N$ derived in \cref{ODA}. The exponential decay constant for the maximum $V_\text{dg}^{i}$ is slightly smaller, around  $0.51\times \ln 2$ for PBC and $0.30\times \ln 2$ for OBC.  In contrast, for the integrable system, the exponential decay constant for the average distinguishability is minimal, about $0.22\times\ln 2$ for PBC and $0.12\times\ln 2$ for OBC. The maximum $V_\text{dg}^{i}$ seems to have almost no decay with $N$ for both boundary conditions. As shown in \cref{FIGs3}(c-d), the difference in off-diagonal elements between integrable and non-integrable systems is small. Consistent with the prediction of \cref{ODA}, the average off-diagonal elements are strictly suppressed exponentially, with an exponential decay constant of $1.00\times\ln 2$ in both cases, regardless of integrability. However, the exponential decay in the average off-diagonal elements does not imply that the maximum values of these elements will also decrease exponentially. As shown in \cref{FIGs3}(d), although the maximum off-diagonal values do decrease with $N$ from the fit: Under PBC, about $0.34\times\ln 2$ for the non-integrable system and about $0.17\times\ln 2$ for the integrable system, they do not exhibit a clear pattern. Under PBC, there is a clear distinction between the integrable and non-integrable cases, and the exponential decay behavior is more pronounced. The exponential decay constant is about $0.78\times\ln 2$ for the non-integrable system and about $0.11\times\ln 2$ for the integrable system.

Additionally, the average off-diagonal elements of the integrable system in \cref{FIGs3}(c) are always slightly smaller than those of the corresponding non-integrable system. This observation is consistent with the fact that the average diagonal elements of the integrable system in \cref{FIGs3}(a) are always larger than the average diagonal elements of the corresponding non-integrable system and the trade-off relationship between the diagonal and off-diagonal elements described in \cref{LVVCAN}.

Moreover, through comparative analysis of PBC and OBC, we observe distinct thermalization characteristics in non-integrable systems. Under OBC configurations, the maximum values of off-diagonal elements in the ETH framework exhibit pronounced exponential decay behavior, a phenomenon likely attributable to the breaking of energy degeneracy. Furthermore, both the average and maximum diagonal elements of ETH demonstrate universally larger magnitudes in OBC systems compared to their PBC counterparts. This can be explained by the absence of the boundary interaction term $-\sigma_z^1 \otimes \sigma_z^{N} $, which impedes efficient information propagation through the system. The resulting suppression of information diffusion creates additional barriers to thermalization, making the equilibration process more challenging in OBC configurations compared to systems with periodic boundaries.

\subsection{The ETH under the average observable}\label{ETHAO}
\begin{figure*}
    \centering
    \subfigure[]{\includegraphics[width=0.48\textwidth]{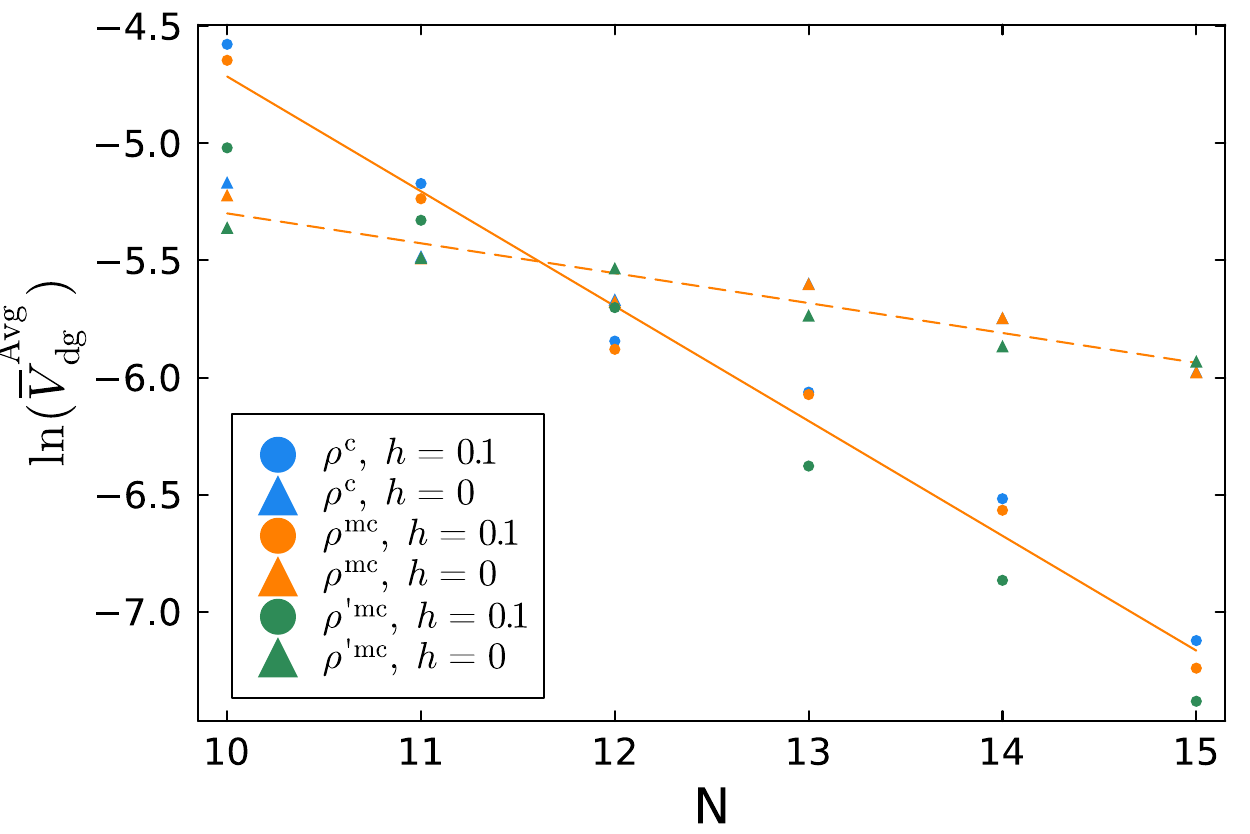}} 
    \subfigure[]{\includegraphics[width=0.48\textwidth]{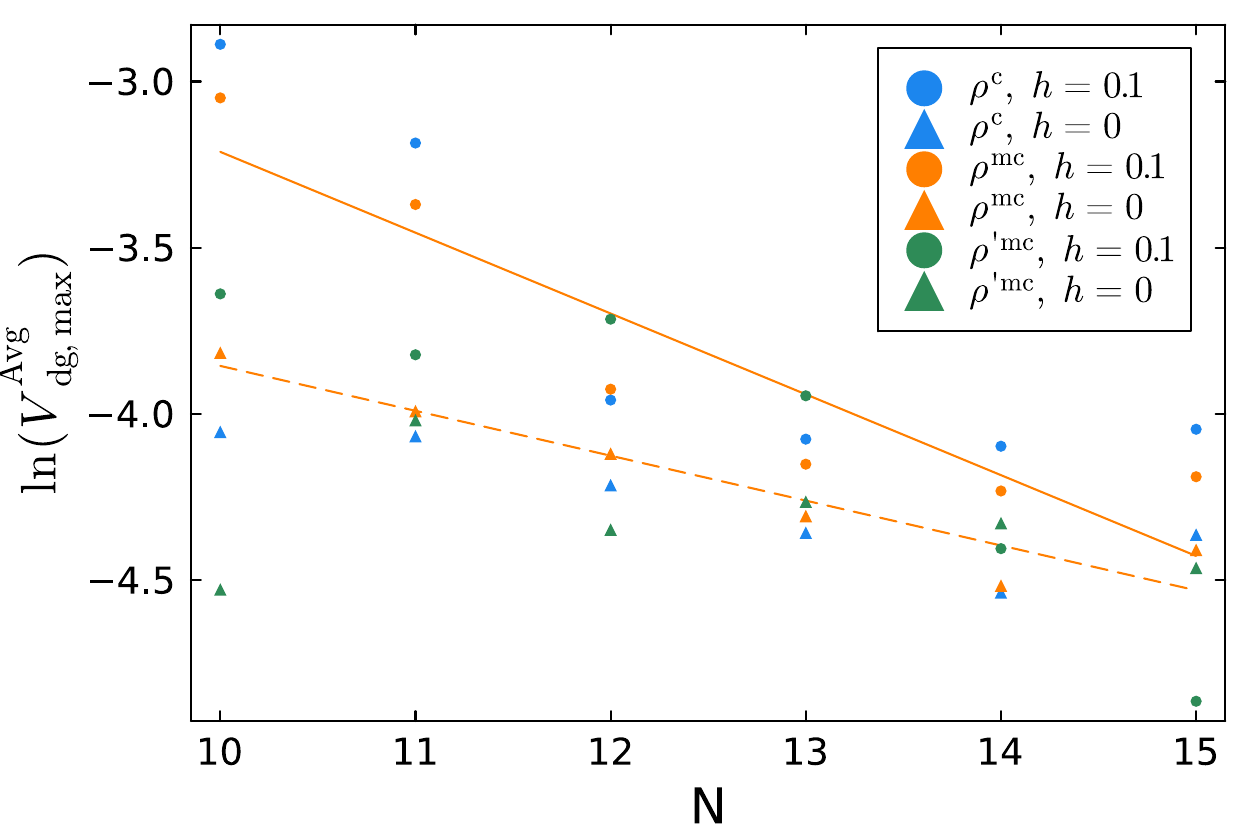}} 
    \subfigure[]{\includegraphics[width=0.48\textwidth]{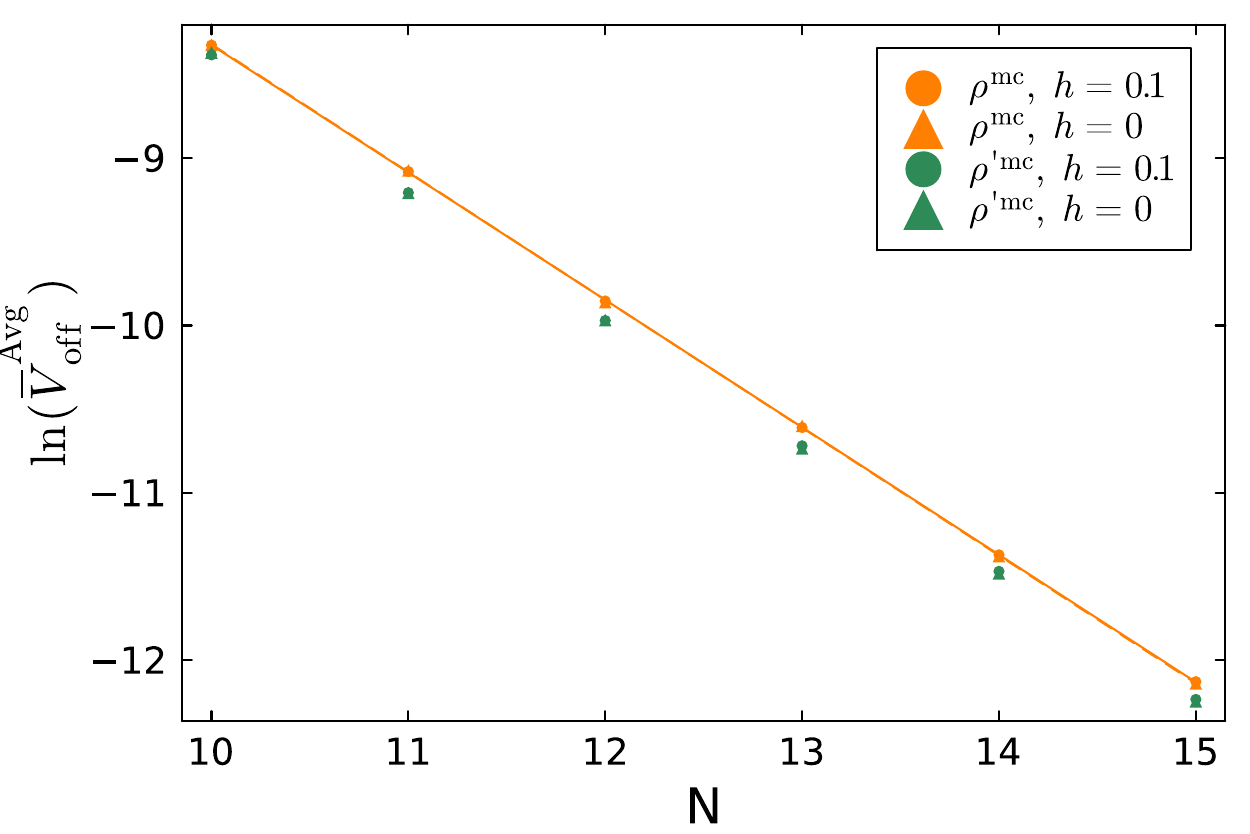}} 
    \subfigure[]{\includegraphics[width=0.48\textwidth]{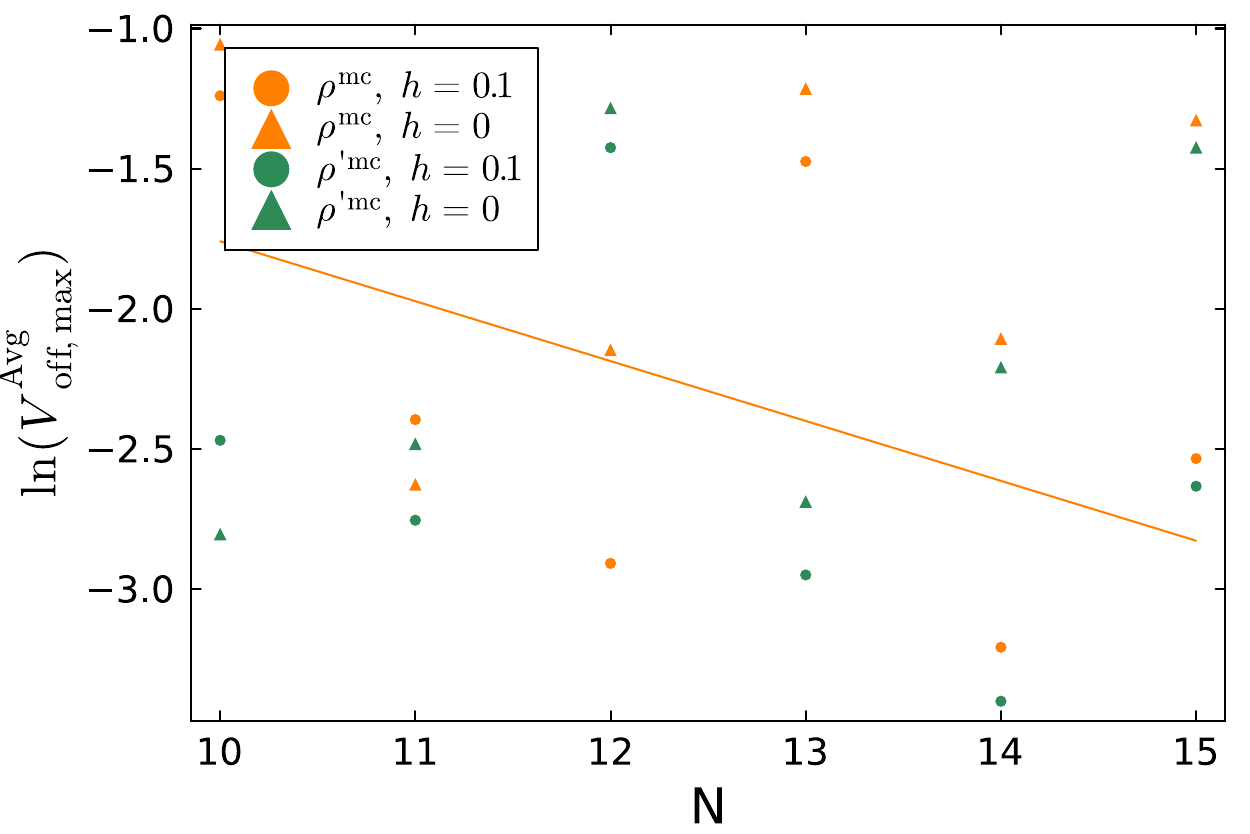}} 
    \caption{Parameters: $g = 1.05$,  $N_{B_1}=1$. Linear fits are shown exclusively for the energy shell $\mathcal{M}_{E(\beta),\Delta}$ with $\rho_B=\rho^{\text{mc}}$. The exponential decay constant is estimated using this linear fit. (a) The average diagonal elements $\overline{V}^\text{Avg}_\text{dg}$ versus $N$. Exponential decay constant are  $0.71\times\ln 2$ (standard error $0.05\times\ln 2$) for the non-integrable system and $0.18\times\ln 2$ (standard error $0.03\times\ln 2$) for the integrable system.  (b) The maximum diagonal element $ V^\text{Avg}_{\text{dg},\max}$ versus $N$. Exponential decay constants are $0.35\times\ln 2$ (standard error $0.07\times\ln 2$) for the non-integrable system and $0.19\times\ln 2$ (standard error $0.03\times\ln 2$) for the integrable system. (c) The average off-diagonal elements $\overline{V}^\text{Avg}_\text{off}$ versus $N$.  Both non-integrable and integrable systems exhibit similar exponential decay constants of $1.1\times\ln 2$  with standard errors $<0.01\times\ln 2$. (d) The maximum diagonal element $ V^\text{Avg}_{\text{off},\max}$ versus $N$. The non-integrable system shows a decay constant of $0.31\times\ln 2$ (standard error $0.26\times\ln 2$). For the integrable case, the decay rate is statistically insignificant (fitted value non-positive due to strong fluctuations) and thus omitted. }
    \label{FIGs4}
\end{figure*}

\begin{figure}
    \centering
    \includegraphics[width=0.48\textwidth]{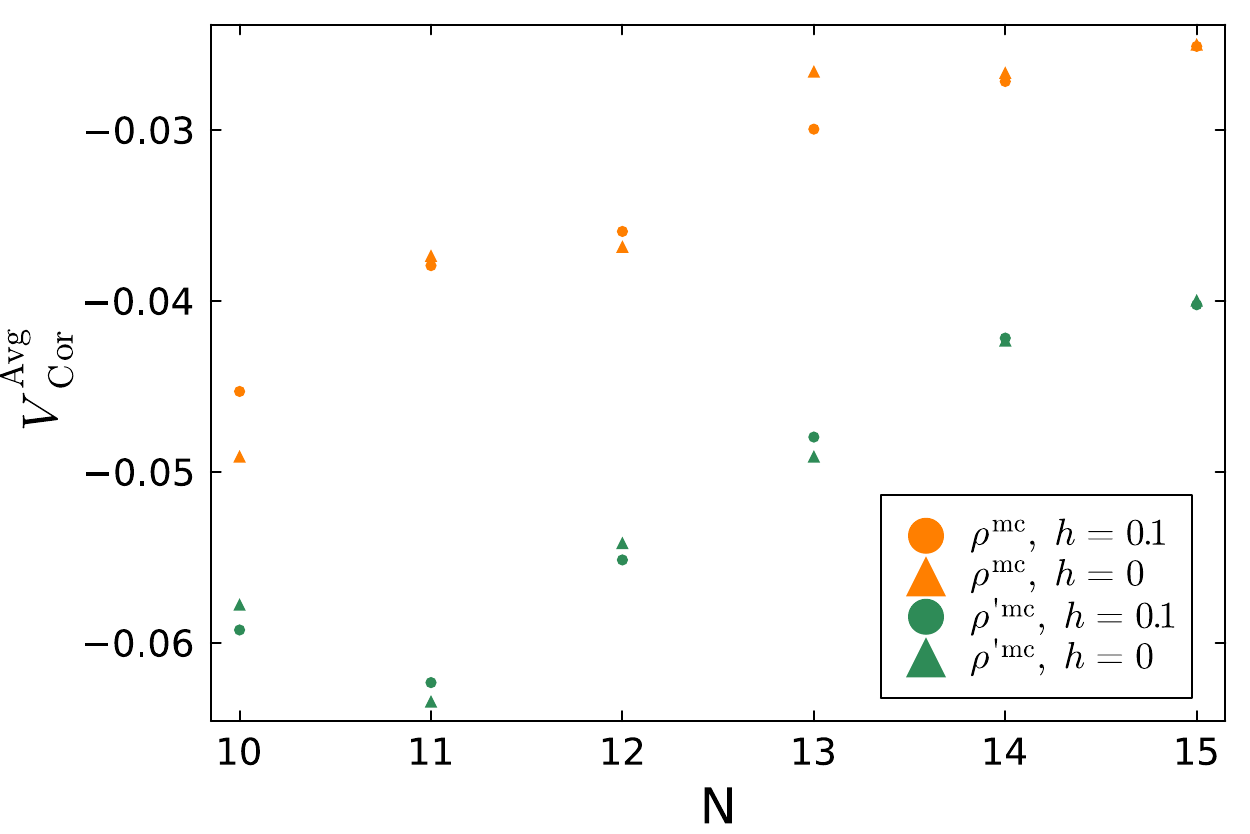}
    \caption{The correlation term $ V^{\text{Avg}}_{\text{Cor}}$ versus $N$. Parameters: $g =1.05$,  $N_{B_1}=1$. }\label{FIG5}
\end{figure}

Now, let us explore the strong and weak ETH under the average observable. The energy shells and ensembles are consistent with those in the previous section. Given that the ensembles used here are all translation-invariant, we simply replace  $V(\rho_B,O^{B_1}_{ij})$ with $ V(\rho_B,\overline{O}^{B_1}_{ij})$ to derive the corresponding $\overline{V}^\text{Avg}_\text{(off-)dg}$ and $V^\text{Avg}_{\text{(off-)dg},\max}$ from \cref{GavgLV,LDGM,LODGM}. According to \cref{avgotdoff,Gl2LoCo}, the local term is obtained by multiplying $\Tr(\rho_{B_1}^{-1})-1$ from the previous section by a factor of $1/\mathcal{C}$.  In the case of the average observable, the spatial correlation of the ensemble states, represented by $  V^{\text{Avg}}_{\text{Cor}}$, also needs to be considered. As shown in \cref{FIG5}, the correlation term is negative, which further lowers the upper bound of the average (off-)diagonal elements. Additionally, the correlation terms vary significantly depending on the energy center, as the ensemble $\rho^\text{mc}$  with energy near zero exhibits weaker correlations. This difference also explains why the two ensembles in \cref{FIGs4}(c) are more different from those in \cref{FIGs3}(c): although the local states of the two ensembles are very similar, their correlation terms are different, which will cause the difference between the two ensembles in \cref{FIGs4}(c) according to \cref{Gl2LoCo}. However, according to \cref{LVVCAN}, the difference in correlation terms cannot be reflected, resulting in almost no difference between the two ensembles in \cref{FIGs3}(c).

As shown in \cref{FIGs4}(a-d), the significant difference in the exponential decay constants of the diagonal elements between integrable and non-integrable systems remains evident. Due to the additional attenuation introduced by the factor $1/\mathcal{C}$ in \cref{Gl2LoCo}, the average of the off-diagonal elements is exponentially suppressed, with an exponential decay constant even larger than $\ln 2$.  The maximum values of the off-diagonal elements still show no clear pattern.  However, in \cref{FIGs4}(c), the average off-diagonal elements of the non-integrable system are no longer consistently smaller than those of the corresponding integrable system. This observation remains  consistent with the trade-off relationship between the diagonal and off-diagonal elements given by \cref{Gl2LoCo}, as \cref{FIGs4}(a) shows that the average diagonal elements of the integrable system are now longer always larger than those of the non-integrable system.

\end{CJK*}

\end{document}